%% file: main.tex
\documentclass[a4paper,10pt,fleqn,usenames,dvipsnames]{article}

\input{preamble}

\usepackage[nonatbib,preprint]{neurips}

\title{Decorrelation using Optimal Transport}
\author{%
  Malte Algren \\
  University of Geneva\\
  \texttt{malte.algren@unige.ch} \\
  \And
  John Andrew Raine \\
  University of Geneva\\
  \texttt{john.raine@unige.ch} \\
  \And
  Tobias Golling \\
  University of Geneva\\
  \texttt{tobias.golling@unige.ch} \\
}

\begin{document}
\maketitle

\begin{abstract}
  Being able to decorrelate a feature space from protected attributes is an area of active research and study in ethics, fairness, and also natural sciences.
  We introduce a novel decorrelation method using Convex Neural Optimal Transport Solvers (Cnots) that is able to decorrelate a continuous feature space against protected attributes with optimal transport.
  We demonstrate how well it performs in the context of jet classification in high energy physics, where classifier scores are desired to be decorrelated from the mass of a jet.
  The decorrelation achieved in binary classification approaches the levels achieved by the state-of-the-art using conditional normalising flows.
  When moving to multiclass outputs the optimal transport approach performs significantly better than the state-of-the-art, suggesting substantial gains at decorrelating multidimensional feature spaces.
\end{abstract}

\input{includes/introduction}

\input{includes/methods}
\input{includes/dataset}

\input{includes/results_1d}

\input{includes/results_3d}

\input{includes/conclusion}

\input{includes/acknowledgement}

\phantomsection
\addcontentsline{toc}{chapter}{References}
\printbibliography[title=References]
\input{includes/appendix.tex}

\end{document}

%% file: preamble.tex
\usepackage{lmodern}
\usepackage[T1]{fontenc}
\usepackage{amsmath}
\usepackage{amssymb}

\usepackage{url}
\usepackage{longtable}
\usepackage{multirow}
\usepackage{caption}
\usepackage[labelformat=simple]{subcaption}

\usepackage[nodayofweek]{datetime}
\usepackage{enumerate}
\usepackage{xspace}
\usepackage{xcolor}
\usepackage{graphicx}
\usepackage[pdftex,hidelinks]{hyperref}
\usepackage{bookmark}

\usepackage{float}
\usepackage[capitalise]{cleveref}

\usepackage[backend=biber,giveninits=true,doi=false,sorting=none,style=numeric-comp]{biblatex}
\DeclareFieldFormat[article]{citetitle}{\textit{#1}\isdot}
\DeclareFieldFormat[article]{title}{\textit{#1}\isdot}
\DeclareFieldFormat[unpublished]{citetitle}{\textit{#1}\isdot}
\DeclareFieldFormat[unpublished]{title}{\textit{#1}\isdot}
\DeclareFieldFormat[inproceedings]{citetitle}{\textit{#1}\isdot}
\DeclareFieldFormat[inproceedings]{title}{\textit{#1}\isdot}

\DeclareNameAlias{author}{last-first}

%% file: includes/introduction.tex
\section{Introduction} \label{sec:intro}
AI-powered decision-making has become a large part of automated systems in banks, advertising, and healthcare, to name a few.
This has resulted in increased awareness surrounding the fairness and biases of the decision models.
Due to the nature of many datasets, biases towards protected attributes like gender and race in data result in biased models.
These biases are not only causes for concern in terms of fairness and ethics, but are also relevant to research in natural sciences, where correlations to protected variables are fundamental in nature, but can lead to undesirable effects in statistical analyses.

In High Energy Physics (HEP), classifiers are commonly used to separate different signal processes from background processes, in both event classification as well as object identification.
One area which has seen a great deal of development is jet tagging, in particular identifying top quark initiated jets from the dominant QCD background of light quarks and gluons (see Ref.~\cite{Kasieczka:2019dbj} for a comprehensive comparison of techniques). 
Identifying the origin of jets is not restricted to supervised classification, with anomaly detection being another area of active development in the hunt for physics beyond the standard model~\cite{dark_machines}. 
Here unsupervised or semisupervised classifiers are used to identify jets which may originate from new physics particles of unknown mass.

In the case of jet tagging, the desired accuracy of the classifier should be independent of the invariant mass of the jet, and instead exploit the differences in the underlying structure of the jet, known as jet "substructure".
However, while background processes often follow an exponentially decaying invariant mass distribution, the invariant mass of the signal processes is localised within some region of the mass spectrum.
This overdensity of signal on the mass spectrum and correlations between the substructure of a jet and its invariant mass will lead to the classifier scores being correlated to the invariant mass.
Several techniques have been developed which aim to decorrelate the scores of a classifier from the invariant mass of a jet~\cite{mode, disco, sam}.
These include methods that are employed during training as a means of regularisation, as well as post-training corrections.

In this work we introduce a new method for decorrelation using Convex Neural Optimal Transport Solvers (Cnots).
Inspired by Ref.~\cite{sam}, which uses normalising flows to learn the monotonic transformation $T(\cdot|c)$ in 1D, given some protected attributes $c$, we propose to use the gradient of a convex neural network~\cite{ICNN} for $T(\cdot|c)$, that by definition is monotonic in $\mathbb{R}^N$. We follow the case of conditional optimal transport studied in Refs.~\cite{gj_chris, bunne2023supervised}.
We use convex neural networks to solve the Kantorovich dual formulation and find the optimal transport (OT)~\cite{villani} between correlated scores and decorrelated ones. Thereby, learning a monotonic transformation $T(\cdot|c)$ between the two spaces, that minimises the Wasserstein distance between them.






%% file: includes/methods.tex
\section{Methods}\label{sec:methods}
\subsection{Dual formulation of optimal transport}

Let $P(y,c)$ and $Q(x,c)$ be two continuous densities, where $x$ and $y$ are coordinates in an $\mathbb{R}^N$ space which follows, different distributions correlated to some latent conditional property $c$.
The optimal transport between these two densities is an optimisation problem over possible transportation maps, $T$, where 
\begin{equation}
    T^* = \inf_{T: T_\#(Q)=P} \frac{1}{2}\mathbb{E}||x-T(x|c)||^2.
    \label{eq:primary_problem}
\end{equation}
Here, $T_\#$ represents possible transports from $Q$ to $P$, $T^*$ is the optimal transport between $Q$ to $P$ and $\mathbb{E}$ is the expectation value over $Q$.


The problem can be formulated with a general cost or distance measure $d(x,y)$, which here is chosen to be the squared Euclidean distance $d(x,y) = ||x-T(x)||^2$.
For this cost function the optimal transport map is unique when $Q$ is continuous~\cite{brenier}.

It is possible to reformulate the primary problem in Eq.~\ref{eq:primary_problem} as a dual formulation following Ref.~\cite{villani} as
\begin{equation*}
    \mathbb{W}^2_2(P,Q) = \sup_{f(y,c)+g(x,c)\leq \frac{1}{2} ||x-y||_2^2} \mathbb{E}(f(y,c)) + \mathbb{E}(g(x,c)), 
\end{equation*}
where $\mathbb{W}^2_2$ is the Wasserstein-2 distance. Here both $f$ and $g$ are functions constrained by
\begin{equation*}
f(y,c)+g(x,c)\leq \frac{1}{2} ||x-y||_2^2.
\end{equation*}
By requiring $f$ and $g$ to be convex functions, this can be rewritten as
\begin{flalign}
    \mathbb{W}^2_2(P,Q) =& \, \mathcal{C}(x,y) 
    +\sup_{f(y,c)\in cvx(y)}\inf_{g(x,c)\in cvx(x)} f(\nabla g(x,c),c)-\langle x,\nabla g(x,c) \rangle - f(y,c), 
    \label{eq:loss_function}
\end{flalign}
where $\mathcal{C}(x,y) = \frac{1}{2}(x^2+y^2)$~\cite{original_ot_NN,villani}. Here both $f$ and $g$ are convex in $x$ and $y$, respectively, but not in $c$.
Under this formulation, the optimal transport map becomes
\begin{equation*}
T_\#(x,c)=\nabla_x g(x,c)=P,
\end{equation*}
the gradient of the convex function $g$ with respect to its inputs $Q$, which, by definition, is monotonic in $x$ for any given $c$.


The monotonic transformation $T_\#$ ensures order preservation in $\mathbb{R}^N$ by definition. For most generative models, this is not important, however, in classification tasks the ordering of jets are important and if not preserved, performance might be lost.
This restriction assists the convergence towards the optimal transport map, which is also order preserving.
One of the benefits of using $\mathbb{W}^2_2$ in comparison to divergences such as the Kullback-Leibler~(KL) divergence is that it only requires samples from the input and base distributions, as opposed to the probability densities.
Furthermore, it is well defined for all values whereas the KL divergence is not. 

\subsubsection*{Convex neural networks}

\begin{figure*}[htpb]
    \centering
    \includegraphics[width=0.7\textwidth]{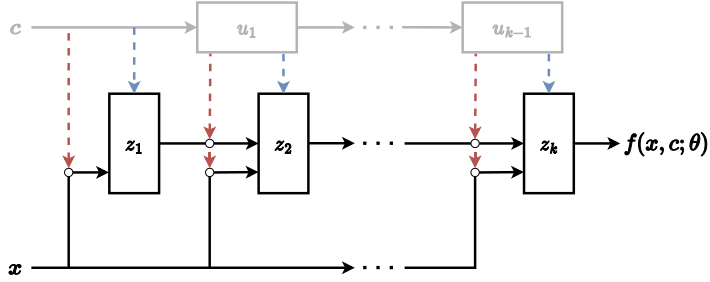}
    \caption{The architecture of the PICNN which parameterises a conditional convex function $f(x,c;\theta)$, that is convex in $x$ but not in $c$. $\theta$ are the trainable parameters of the PICNN network. $\circ$ symbols indicate Hadamard products.
    The convex part of the network is indicated by the black lines and consists of recursive blocks $z_{1,...,k}$, that is defined in Eq.~\ref{eq:cvx_part}. The grey lines are the non-convex component consisting of $u_{1,...,(k-1)}$, defined in Eq.~\ref{eq:noncvx_part}. The architecture reduces to an ICNN when removing the non-convex part. }
    \label{fig:PICNN}
\end{figure*}

In Ref~\cite{original_ot_NN}, it is shown that Eq.~\ref{eq:loss_function} can be solved by parameterising the two convex functions with two Input-Convex neural networks (ICNN) \cite{ICNN} and thus learning the optimal transport as $\nabla_x g(Q;\theta)=P$ and $\nabla_y f(P;\theta)=Q$, where $\theta$ is the trainable parameters of the network. Partially input convex neural networks (PICNN)~\cite{ICNN} extend ICNNs to take a conditional vector $c$ in addition to the input vector $x$ in order to represent a conditional convex scalar function $f(x,c;\theta)$.
In this case $f(x,c;\theta)$ does not need to be convex with respect to $c$.
The general architecture of the PICNNs can be seen in Fig.~\ref{fig:PICNN}, whereby removing the non-convex part will reduce it to the ICNN.


Following the same notation as in Ref~\cite{ICNN}, the PICNN is defined recursively in $z_k$ and $u_k$, where $k$ is the number of layers ranging from  $i=0,1,...,k$, and $W$ and $b$ are learnable parameters.
The recursive non-convex layers follow a simple feed-forward network
\begin{align}
    u_{i+1} = \widetilde{g_i}(\widetilde{W}_i u_i+\widetilde{b}_i),
    \label{eq:noncvx_part}
\end{align}
with $u_0 = c$ and $\widetilde{g_i}$ is the activation function.
These non-convex layers have no restrictions on allowed transformations and operations, as long as they are differentiable.
The convex layers are defined as
\begin{align}
    z_{i+1} = g_i \bigl( W_{i, SP}^{z} \bigl( z_i \circ [W_i^{zu}u_i+b^{z}]_{SP} \bigr) 
    + W_i^{y} \bigl( y \circ (W_i^{yu} u_i + b_i^{y}) \bigr) + W_i^u u_i + b_i \bigr),
    \label{eq:cvx_part}
\end{align}
where $z_0=0$.
$W_{i, SP}^{z}$ are restricted to be strictly positive by passing the weights through a softplus transformation, and the activation functions $g_i$ are required to be convex and non-decreasing. Lastly, the output is defined as
\begin{equation}
    f(x,c;\theta) = \frac{w_{0, SP}(u_k)}{2}x^2 + w_{1, SP}(u_k) z_k,
\end{equation}
such that the transport function given by
\begin{equation}
    \nabla_xf(x,c;\theta) = w_{0, SP}(u_k) \cdot x + w_{1, SP}(u_k) \nabla_x z_k
\end{equation}
makes perturbations around the identity function easily accessible and the transport at initialisation for random weights $\theta$.

\subsection*{Convex neural networks for decorrelation}
In order to use convex neural networks for decorrelation purposes, the aim is to learn a monotonic transformation $T(\cdot|c)$ between the input feature space $Q(\cdot|c)$ and a target feature space $P(\cdot)$.
To break the dependence on $c$, we construct $P$ to be a distribution which is identical for all values $c$.
In this framework, the choice of $P$ is completely free and does not need to follow an analytically defined function.
This is in contrast to normalising flows, which require the probability distribution function of the base density to be analytically calculable.
Alternatively, normalising flows can also use arbitrary PDFs as base distributions \cite{flows4flows}, though we do not study this here.
The simplest way to construct $P(\cdot)$ is to take the distribution $Q(\cdot|c)$ and randomly permute the conditional vector, breaking the correlation.
This should also define a target base distribution which is similar to the input distribution, simplifying the transport function.
Other choices are to choose a uniform distribution or normal distribution, as is done for cflows in Ref.~\cite{sam}.

\subsection{Other exiting methods} \label{sec:other_methods}
\subsubsection*{Conditional normalising flows}
Conditional normalising flows (cflows) \cite{flows_v2, flows_v1} are networks built to be fully invertible, therefore by definition $f$ and $f^{-1}$ exist and are fast to compute. This makes cflows powerful as generative models using the change of variables formula
\begin{equation}
    p_x(x|c) = p_y(f(x,c)) \left| \det \mathcal{J}(f(x,c)) \right| \label{eq:change_of_variables},
\end{equation}
to transform between base density $p_y(x)$ independent in $c$ and some complex density $p_x(x|c)$ conditioned on $c$. 
The objective of the cflow will be to maximise the log-likelihood, which requires the knowledge of the PDF of the base density $p_y(x)$. By construction, the cflow can be inverted $f^{-1}(p_x(x|c))\rightarrow p_y(x)$ to be independent of $c$, producing the desired decorrelated features~\cite{sam}.

For $p_x(x|c)$ in 1D, the cflow transformation can be restricted to be monotone ensuring order preservation, which is important for decorrelating discriminate scores. However, beyond 1D, the monotonicity of the transformations is not guaranteed. 

An invertible PICNN architecture can also be trained like a normalising flow to maximise log-likelihood \cite{cp_flows} and find a monotonic transformation in $\mathbb{R}^N$.
However, this involves calculating the Hessian in the forward pass, which is an expensive procedure and limits the choice of base distribution to analytically defined functions.
Due to the additional complexity, this training scheme is not studied in this work.

\subsubsection*{Decorrelating during training}
A wide range of established methods for the decorrelation of classifier outputs are applied before or during the training of the discriminator.
Planing~\cite{planing} can be applied to the data beforehand, as a form of preprocessing, ensuring that the distribution over the protected variables $c$ follow the same distribution for both the signal and background jets.
An alternative approach is penalising the classifier for producing outputs which are correlated with $c$ during training by adding an additional loss or regularisation term $\mathcal{L}_{corr}$.
Example methods calculate $\mathcal{L}_{corr}$ with adversarial neural networks~\cite{louppe2017learning, PhysRevD.96.074034, Windischhofer:2019ltt}, or with distance measures calculated using distance correlation~\cite{disco} or the moments of the conditional cumulative distributions~\cite{mode}.
The total loss for these methods is given by
\begin{equation}
    \mathcal{L} = \mathcal{L}_{class} (s(x), y)+\alpha \mathcal{L}_{corr}(s(x), c),
\end{equation}
$c$ being the protected attributes, $y$ are the labels and $s(x)$ is the classifer output during training. The decorrelation can then be controlled by a hyperparameter $\alpha$.




Both the optimal transport and cflow decorrelation approaches are applied only after training, or as a means of preprocessing, we restrict comparisons to these two methods.
Furthermore, as shown in Ref.~\cite{sam} for cflows, the OT approach can also be applied in addition to other decorrelation approaches, such as those used during training.

%% file: includes/dataset.tex
\section{Application to jet tagging} \label{sec:studies}


In HEP, multiple studies have been conducted to decorrelate the predictions $s(x|m)$ of binary classifiers from the invariant mass $m$ of a reconstructed object~\cite{ATL-PHYS-PUB-2018-014, sam, mode, disco, planing, louppe2017learning, PhysRevD.96.074034, Windischhofer:2019ltt}.
However, balancing the classification loss and decorrelation loss has proven difficult. The current state-of-the-art approach uses conditional flows to and imposes no restrictions on either the dataset or the architecture of the classifier.




To form a basis of comaprison and evaluate the performance between Cnots and cflows, we look at  decorrelation of classifiers trained to identify the origin of boosted objects at particle colliders like the LHC~\cite{LHC}.
When collisions produce particles with high transverse momentum, their decay products have a smaller opening angle.
In the case where these particles decay to partons, the two resulting hadronic showers, known as jets, start to overlap.
In this instance, their decay products are unable to be resolved individually and instead are reconstructed as a single large jet.
Differences in the underlying structure within the jet can be exploited to predict the initial particle produced in collisions. However, the underlying structure of jets remains strongly correlated to the reconstructed invariant mass of the jet resulting in a biased prediction of the initial particle.
In this work we study the performance of decorrelating classifiers trained to identify boosted jets origination from quarks and gluons (QCD), vector bosons (VB) and top quarks (Top), from the invariant mass of the jet.
\subsection*{Datasets}

To get pure samples of the top quark initiated jets, samples of $t\bar{t}$ jets are produced, in which both the top quarks decay hadronically, ($t\rightarrow W(\rightarrow q\bar{q}^{\prime})b)$.
For a pure sample of vector boson initiated jets, samples of $WZ$ diboson events are generated in which both the $W$ and $Z$ bosons decay to two quarks.
For pure samples of QCD initiated jets, samples of two-to-two processes with a final state of two quarks and or gluons are simulated.
All three samples are generated at a centre of mass energy $\sqrt{s}=$~13~TeV using MadGraph\_aMC@NLO~\cite{MadGraph}~(v3.1.0), with decays of top quarks and $W$ bosons modelled with MadSpin \cite{MadSpin}.
Pythia8~\cite{Pythia}~(v8.243) is used to model the parton shower and hadronisation with the NNPDF2.3LO PDF set~\cite{PartonDFs}.

The detector response is simulated using Delphes~\cite{Delphes}~(v3.4.2) with a parametrisation similar to the ATLAS detector~\cite{ATLAS}.
Jets are reclustered using the anti-$k_t$ clustering algorithm~\cite{AntiKt} with a radius parameter $R=1.0$ using the \texttt{FastJet} package~\cite{FastJet}.
Jets are required to have $p_\mathrm{T}>450$~GeV and $|\eta|<2.5$, with only the jet with the highest $p_\mathrm{T}$ selected from each event.
The minimum $p_\mathrm{T}$ of the leading parton in the hard scatter is optimised for each sample in order to increase the rate of jets passing the selection criteria and in order to produce similar distributions for the jet $p_\mathrm{T}$ across all three jet types.

The relative four momenta~($p_\mathrm{T}^{\text{frac}}$, $\Delta\eta$, $\Delta\phi$, $E^\text{frac}$) of up to the leading 100 constituents, ordered in descending $p_\mathrm{T}$, are stored for each jet, alongside the jet four-momentum vector~($p_\mathrm{T}$, $\eta$, $\phi$, $m$).
Jets with fewer than 100 constituents are zero-padded.
In total, there are 840,000 each of Top, VB, and QCD jets in the training set, and 800,000, 93,000 and 900,000 jets, respectively, for each class in the test set. Only jets with an invariant mass between 20 and 450 GeV are selected for decorrelation.

\subsection*{Classifiers}

A multiclass classifier (mDNN) is trained to predict the probabilities of a jet originating from each jet type, $p_{QCD}, p_{T}$ and $p_{VB}$.
The mDNN is constructed using the Particle-Transformer architecture from Ref.~\cite{transformer}, and is trained using the constituents of the jets. 

Whilst the multi dimensional scores from the mDNN are correlated to mass, visualising the correlated scores in $\mathbb{R}^3$ is difficult. 
Thus, for visualisation purposes, we project the 3D scores down to 1D distributions following the Neyman Peason lemma to create three discriminators $\mathcal{D}$

\begin{align}
    \mathcal{D}_\mathrm{QCD} = \frac{p_{QCD}}{p_{T}+p_{VB}}, \quad
    \mathcal{D}_\mathrm{T} = \frac{p_{T}}{p_{QCD}+p_{VB}}, \quad
    \mathcal{D}_\mathrm{VB} = \frac{p_{VB}}{p_{T}+p_{QCD}}.
    \label{eq:3d_discriminate}
\end{align}

To evaluate the performance of decorrelating the output of a binary classifier, we use the discriminator scores $\mathcal{D}_\mathrm{VB}$ normalised with a sigmoid transformation.
For the decorrelation of a multiclass output, we decorrelate the joint distribution $p_{QCD}, p_{T}$ and $p_{VB}$.
In the 3D case, the discriminators in Eq.~\ref{eq:3d_discriminate} are used for visualisation.

\begin{figure}[htpb]
    \centering
    \includegraphics[width=1\textwidth]{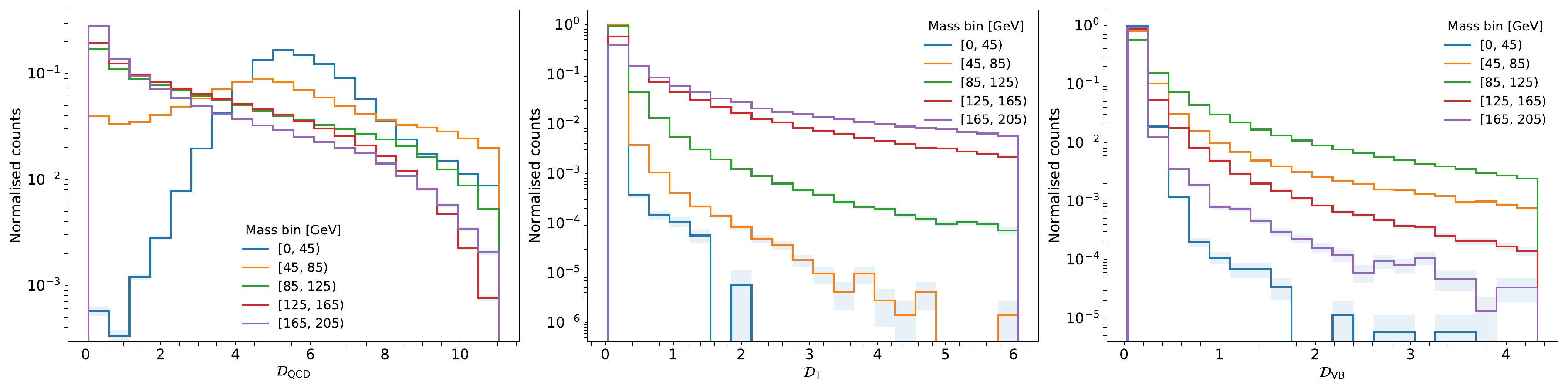}
    \caption{The scores from the Transformer projected down to 1D discriminators using Eq.~\ref{eq:3d_discriminate}.} \label{fig:multicls_prediction_correlated}
\end{figure}

\begin{figure*}[htpb]
    \centering
    \begin{subfigure}[t]{0.33\textwidth}
        \centering
        \includegraphics[width=1\textwidth]{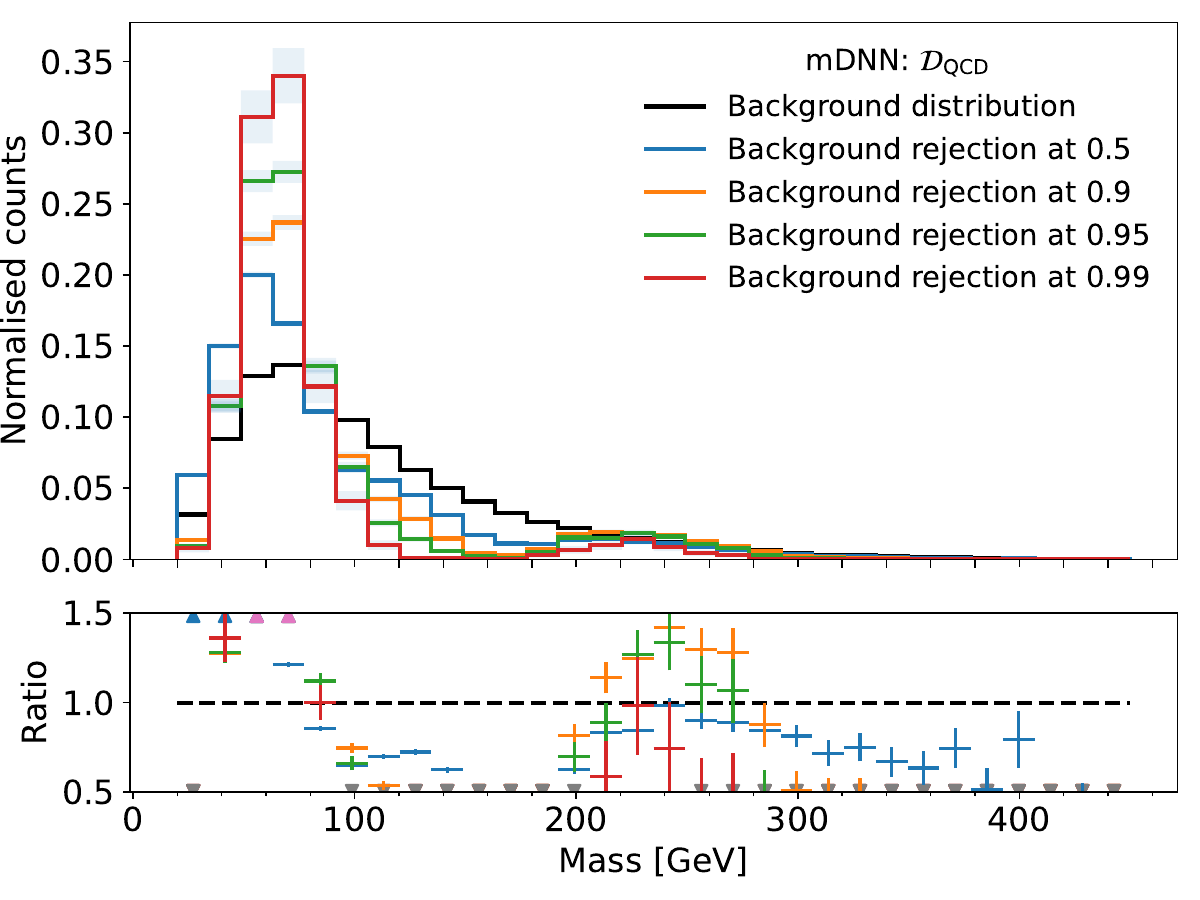}
        \caption{$\mathcal{D}_\mathrm{QCD}$ projection}
    \end{subfigure}%
    \begin{subfigure}[t]{0.33\textwidth}
        \centering
        \includegraphics[width=1\textwidth]{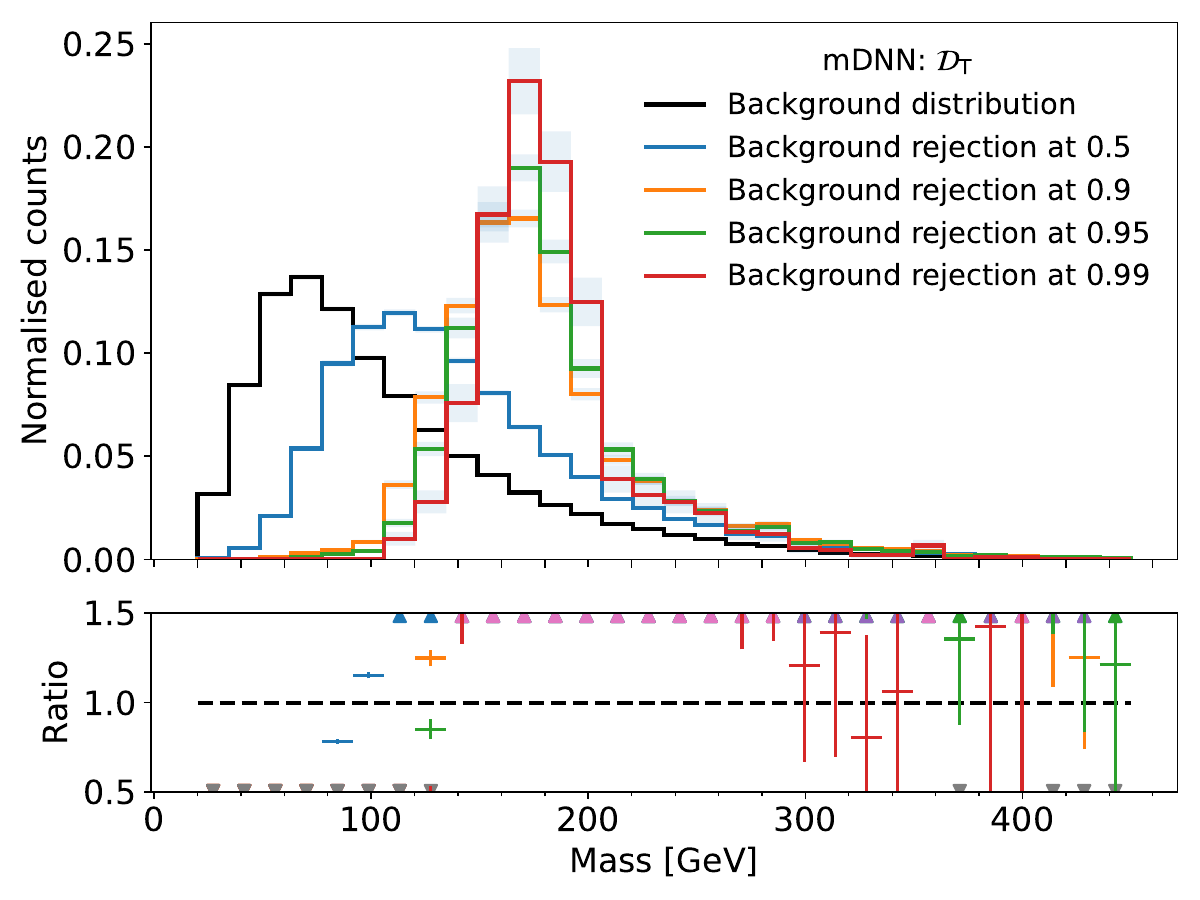}
        \caption{$\mathcal{D}_\mathrm{Top}$ projection}
    \end{subfigure}
    \begin{subfigure}[t]{0.33\textwidth}
        \centering
        \includegraphics[width=1\textwidth]{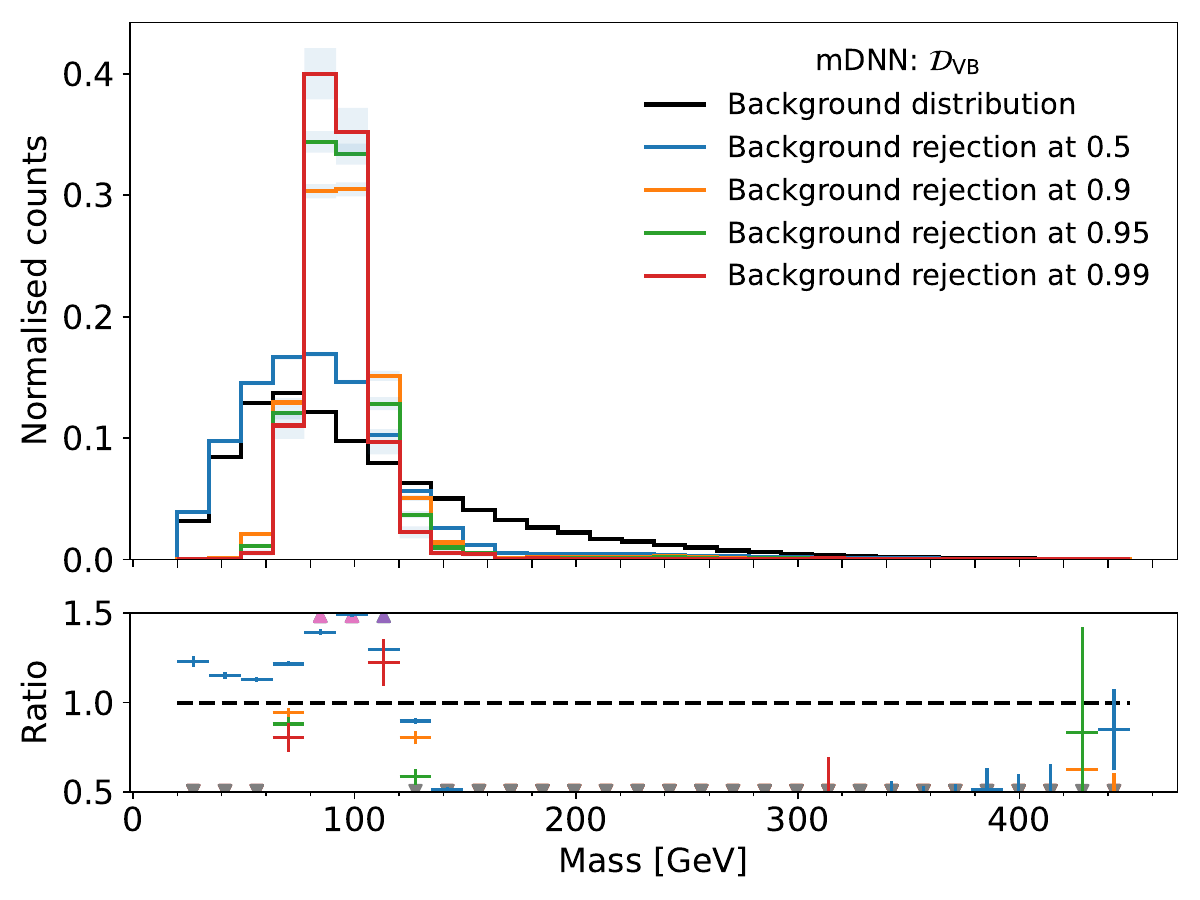}
        \caption{$\mathcal{D}_\mathrm{VB}$ projection}
    \end{subfigure}
    \caption{Mass distribution at background rejections of 50\%, 90\%, 95\% and 99\% for the raw discriminators.}
    \label{fig:3d_original_mass_sculpting}
\end{figure*}

The three discriminators are shown in Fig.~\ref{fig:multicls_prediction_correlated}, where we can see that the scores change as a function of mass.
The mass sculpting is also very apparent when applying a selection on the discriminate scores, as seen in Fig.~\ref{fig:3d_original_mass_sculpting}, where, especially for the $\mathcal{D}_\mathrm{VB}$ and $\mathcal{D}_\mathrm{T}$, the sculpting surrounding the resonance mass is evident. 

%% file: includes/results_1d.tex
\section{Results}
While the decorrelation methods are trained exclusively on QCD jets to ensure that there is no underlying correlation between classes, evaluation includes all classes. 

To evaluate the decorrelation performance, prior decorrelation studies in HEP have used the inverse Jensen-Shannon divergence (1/JSD) between the initial invariant mass distribution and the distribution after applying a selection on the classifier scores. This is an effective measure of the mass sculpting resulting from the classifier output.
As the decorrelation methods are trained on QCD jets, fully decorrelated classifier scores should not sculpt the initial mass distribution after a selection.
In order to estimate an upper bound on performance arising from the statistical variation of the data, we calculate the ideal 1/JSD using bootstraps \cite{bootstrap} on the initial QCD mass distribution without a selection.
To measure the discrimination power after decorrelation, we calculate the signal efficiency for both the Top and VB as a function of background rejection, and calculate how the AUC changes as a function of mass.

\subsection{Binary decorrelation} \label{sec:binary_decorr}
\begin{figure*}[htpb]
    \centering
    \begin{subfigure}[t]{0.48\textwidth}
        \centering
        \includegraphics[width=1\textwidth]{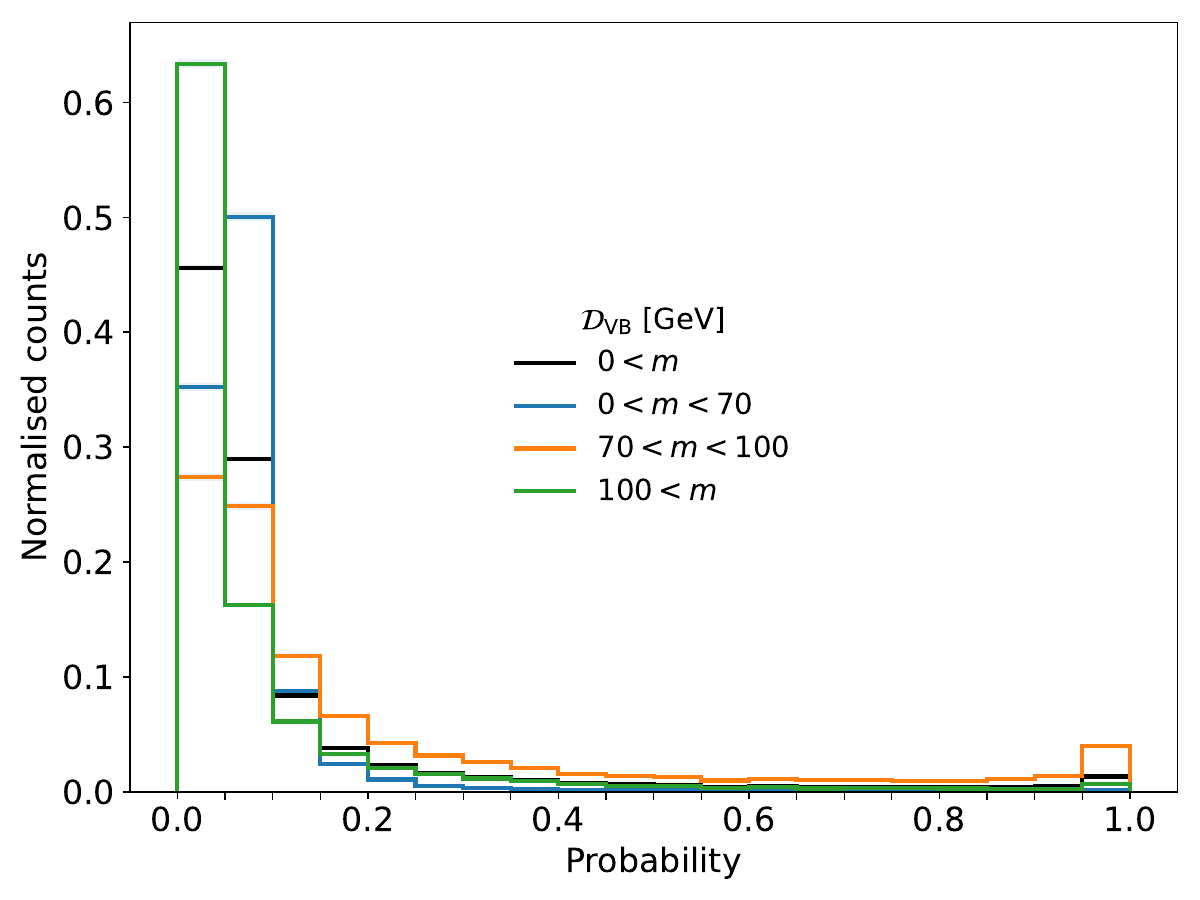}
        \caption{$\mathcal{D}_\mathrm{VB}$ scores in different mass bins. \label{fig:DNN_disc_dist}}
    \end{subfigure}%
    \,
    \begin{subfigure}[t]{0.48\textwidth}
        \centering
        \includegraphics[width=1\textwidth]{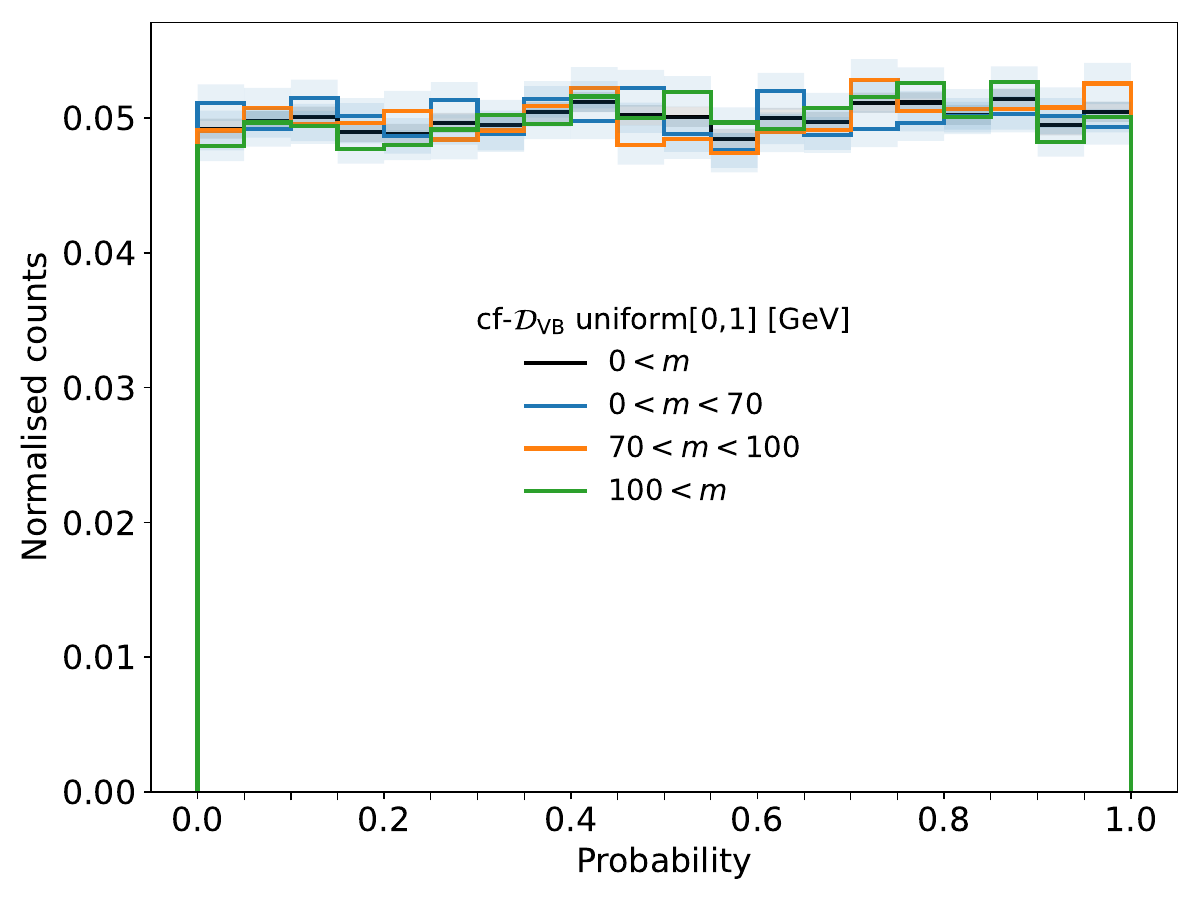}
        \caption{Flow decorrelated $\mathcal{D}_\mathrm{VB}$ scores using a uniform distribution as a base distribution. \label{fig:flow_disc_dist}}
    \end{subfigure}
    \begin{subfigure}[t]{0.48\textwidth}
        \centering
        \includegraphics[width=1\textwidth]{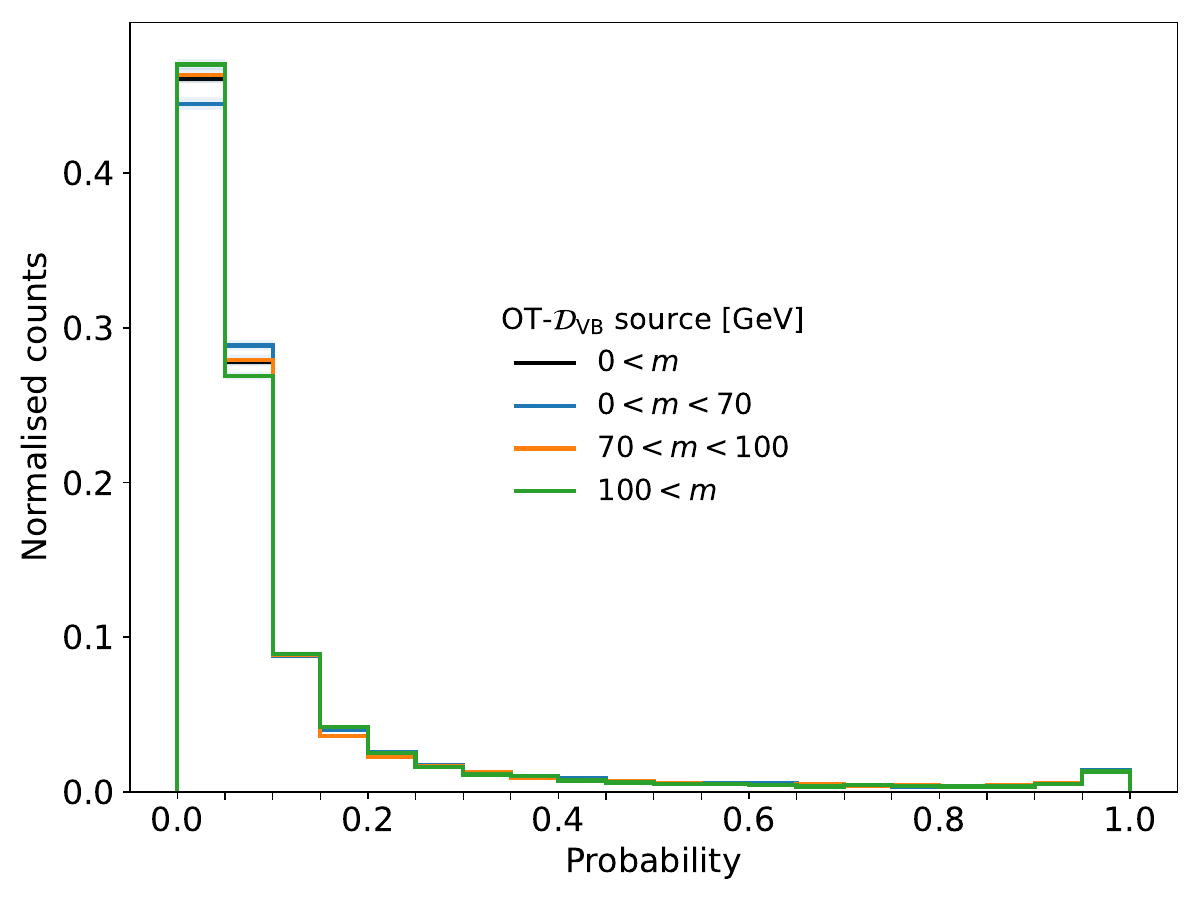}
        \caption{OT decorrelated $\mathcal{D}_\mathrm{VB}$ scores using the original $\mathcal{D}_\mathrm{VB}$ distribution. \label{fig:OT_same_disc_dist}}
    \end{subfigure}
    \,
    \begin{subfigure}[t]{0.48\textwidth}
        \centering
        \includegraphics[width=1\textwidth]{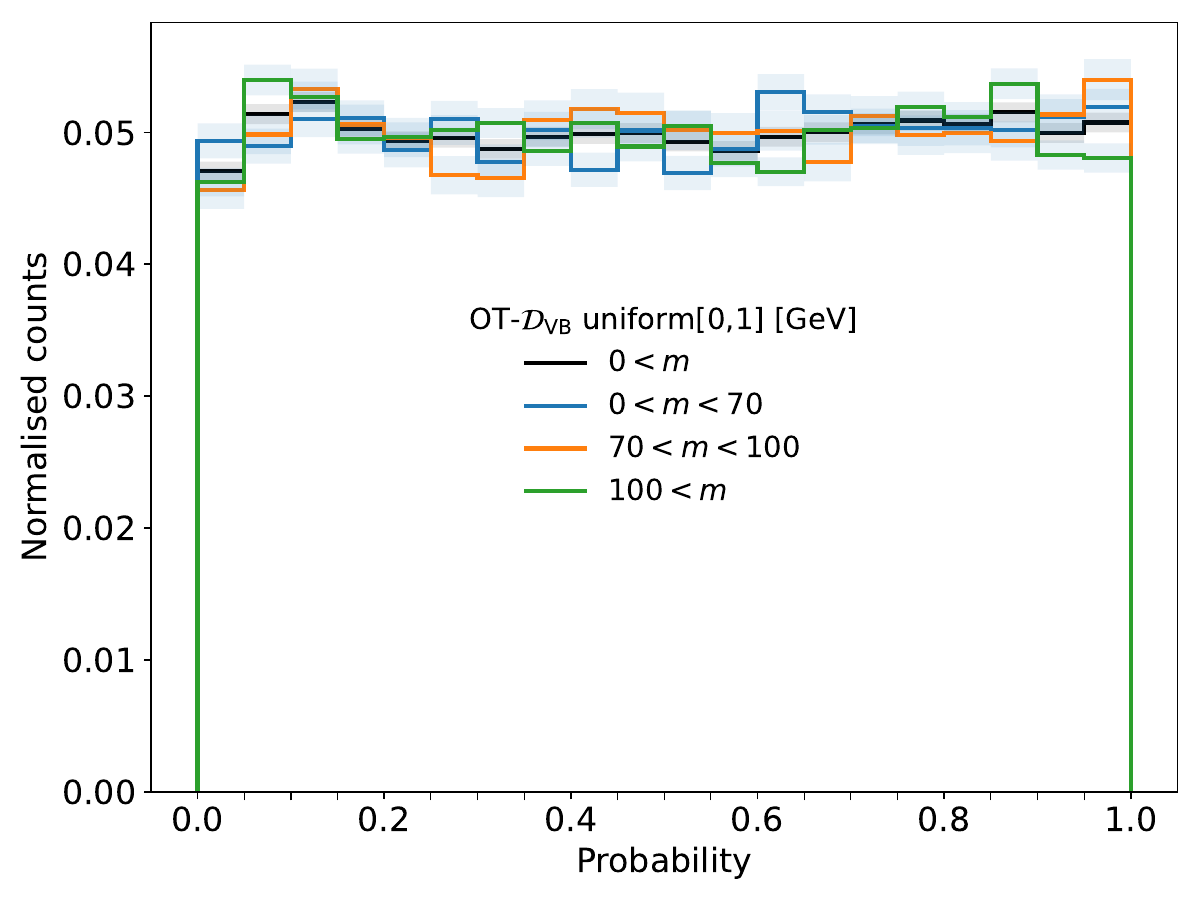}
        \caption{OT decorrelated $\mathcal{D}_\mathrm{VB}$ scores using a uniform distribution as a base distribution. \label{fig:OT_uni_disc_dist}}
    \end{subfigure}
    \caption{The $\mathcal{D}_\mathrm{VB}$ scores in different mass bins. (a) shows the original distribution. In (b), (c) and (d), the scores are decorrelated to different base distributions using the decorrelation methods explained in Sec.~\ref{sec:methods}.} \label{fig:all_disc_dist}
\end{figure*}

The correlated $\mathcal{D}_\mathrm{VB}$ scores are shown in Fig.~\ref{fig:all_disc_dist} as a function of mass, where a dependency on mass is apparent, especially around the W/Z-boson mass.

In Fig.~\ref{fig:all_disc_dist}, we trained the decorrelation methods explained in Sec.~\ref{sec:methods} and applied the learned transformation to decorrelate the $\mathcal{D}_\mathrm{VB}$ scores. 
For both methods, the mass dependency is removed and the distribution of the scores are the same in the four mass bins.
The Cnots method for 1D decorrelation of $\mathcal{D}_\mathrm{VB}$ (OT-$\mathcal{D}_\mathrm{VB}$) is able to decorrelate to an arbitrary base distribution, whereas the cflow method has to evaluate the likelihood of the PDFs. Therefore, we show two possible base distributions for Cnots, one using a uniform distribution, and another using the source distribution as the base distribution. 

\begin{figure}[htpb]
    \centering
    \includegraphics[width=0.5\textwidth]{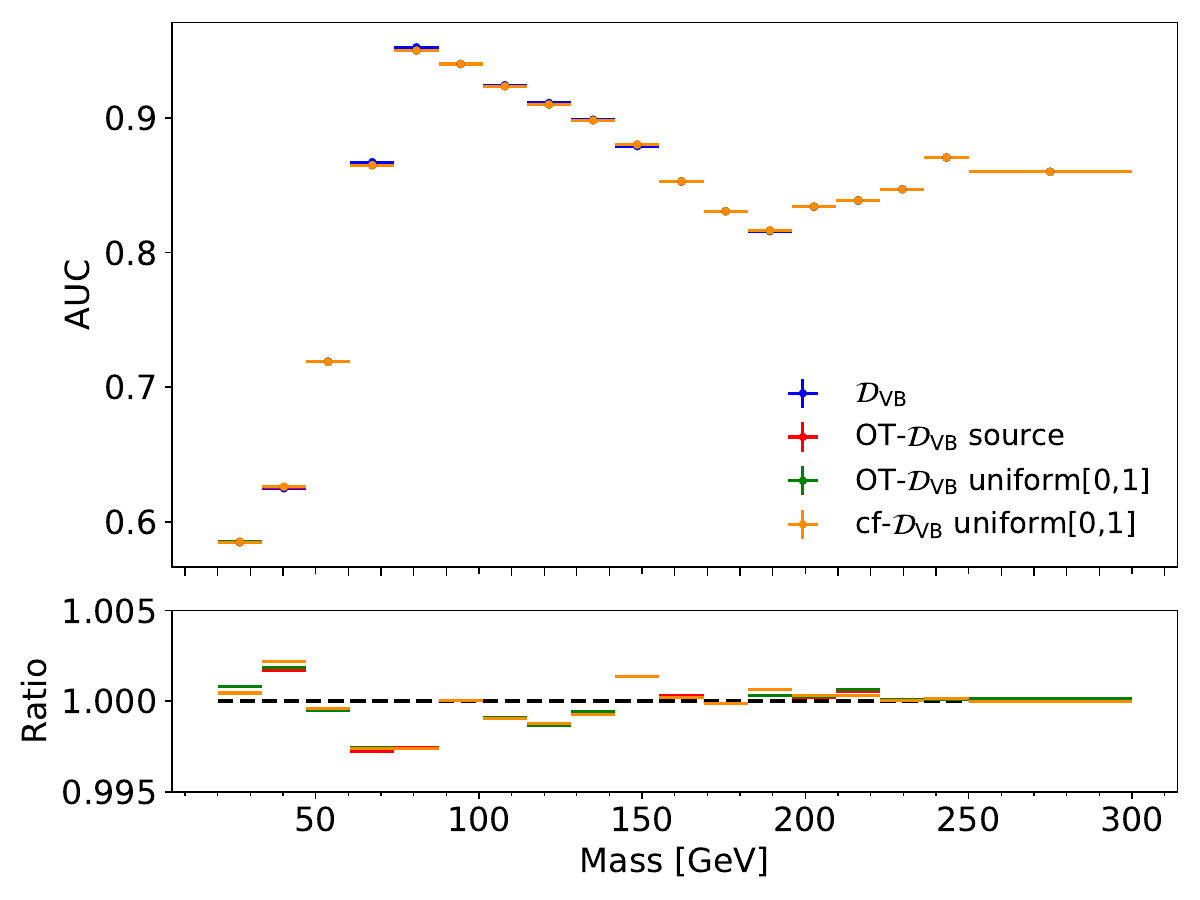}
    \caption{Comparison between the AUC before and after decorrelation as a function of mass.
    The ratio is calculated between the binned AUC values of the decorrelated $\mathcal{D}_\mathrm{VB}$ and the initial discriminator.
    } \label{fig:AUC_mass_1d}
\end{figure}

After decorrelation, the inclusive separation power will degrade, as $\mathcal{D}_\mathrm{VB}$ scores are not able to discriminate using the additional separation power arising from the jet mass.
However, due to the monotonicity of the transformations, the integrated performance over the protected attributes should remain the same. In Fig.~\ref{fig:AUC_mass_1d}, we select jets within narrow mass bins to imitate the integration and calculate the AUC.
We see here that the AUC as a function of mass remains unchanged after decorrelation.

\begin{figure*}[htpb]
    \centering
    \begin{subfigure}[t]{0.49\textwidth}
        \centering
        \includegraphics[width=1\textwidth]{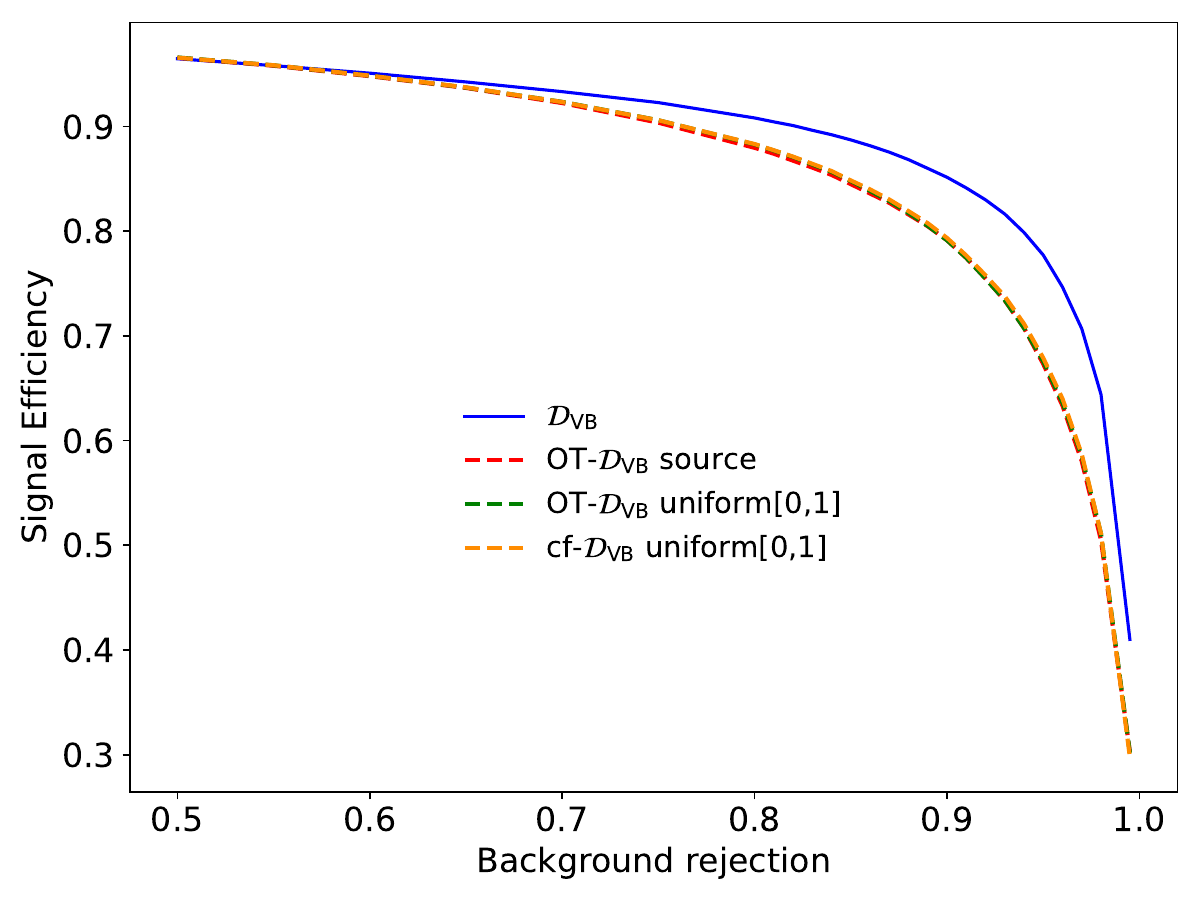}
        \caption{The signal efficiency as a function of background rejection for the $\mathcal{D}_\mathrm{VB}$ scores and the decorrelation methods. \label{fig:sig_eff_1d}}
    \end{subfigure}%
    \,
    \begin{subfigure}[t]{0.49\textwidth}
        \centering
        \includegraphics[width=1\textwidth]{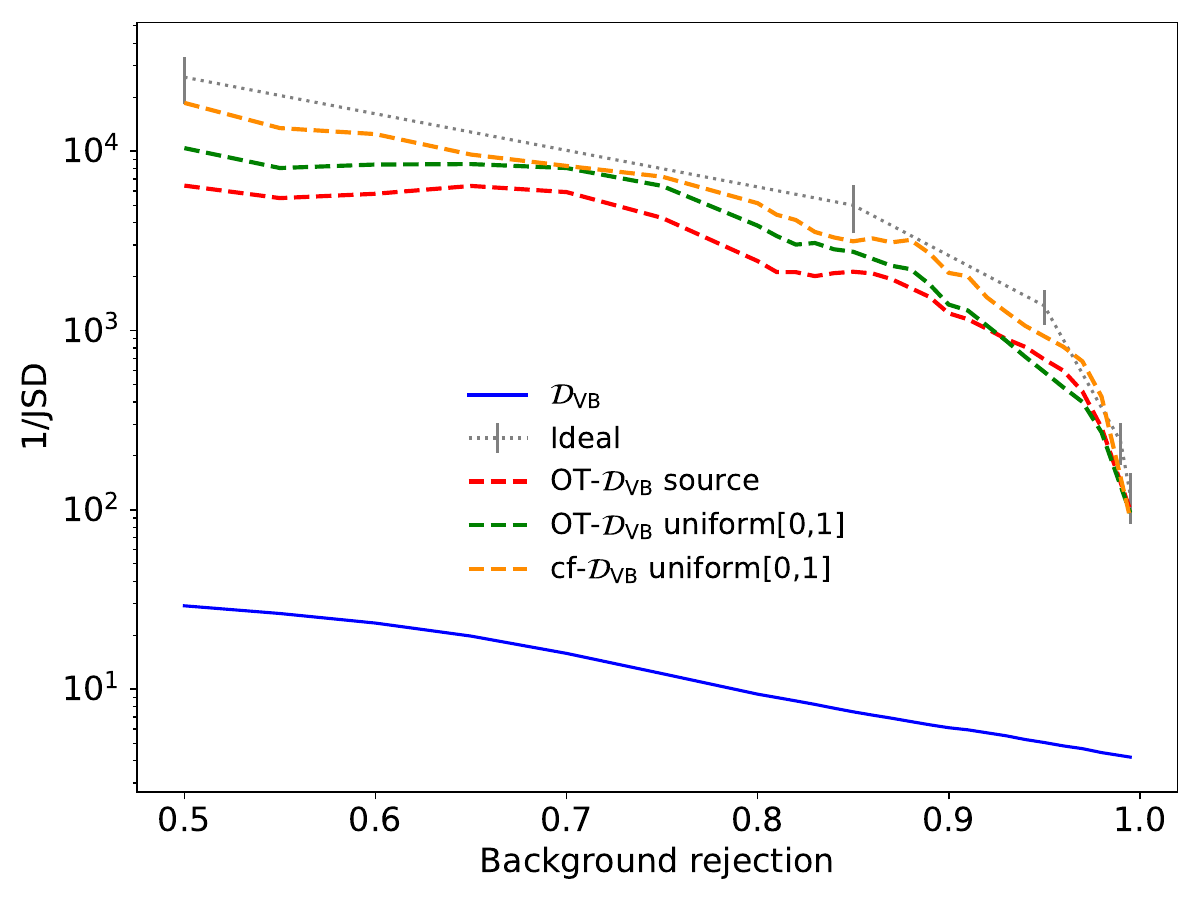}
        \caption{The 1/JSD values as a function of background rejection for the $\mathcal{D}_\mathrm{VB}$ scores and the decorrelation methods. The ideal 1/JSD has been found by bootstrapping the pure background distribution \label{fig:bkg_recj_1d}}
    \end{subfigure}
    \caption{The signal efficiency and 1/JSD are measured at various levels of background rejection. (a) shows the signal efficiency at the different background rejections. (b) shows the background sculpting at the different background rejections. 1/JSD is measured only on QCD jets.}
    \label{fig:decorrelation_performance}
\end{figure*}

To evaluate the decorrelation performance, we measure the 1/JSD at different selections and simultaneously measure the signal efficiency of VB. This is illustrated in Fig.~\ref{fig:decorrelation_performance} for the $\mathcal{D}_\mathrm{VB}$, cf-$\mathcal{D}_\mathrm{VB}$ and OT-$\mathcal{D}_\mathrm{VB}$.
While we see no large variation in the signal efficiency between the decorrelation, the cflow method outperforms the OT methods in the background sculpting. At very high background rejection we see comparable performances, as both methods are within the statistical uncertainty of the ideal background distribution.


%% file: includes/results_3d.tex
\subsection{Multiclass decorrelation}

\begin{figure*}[htpb]
    \centering
    \begin{subfigure}[t]{1\textwidth}
        \centering
        \includegraphics[width=1\textwidth]{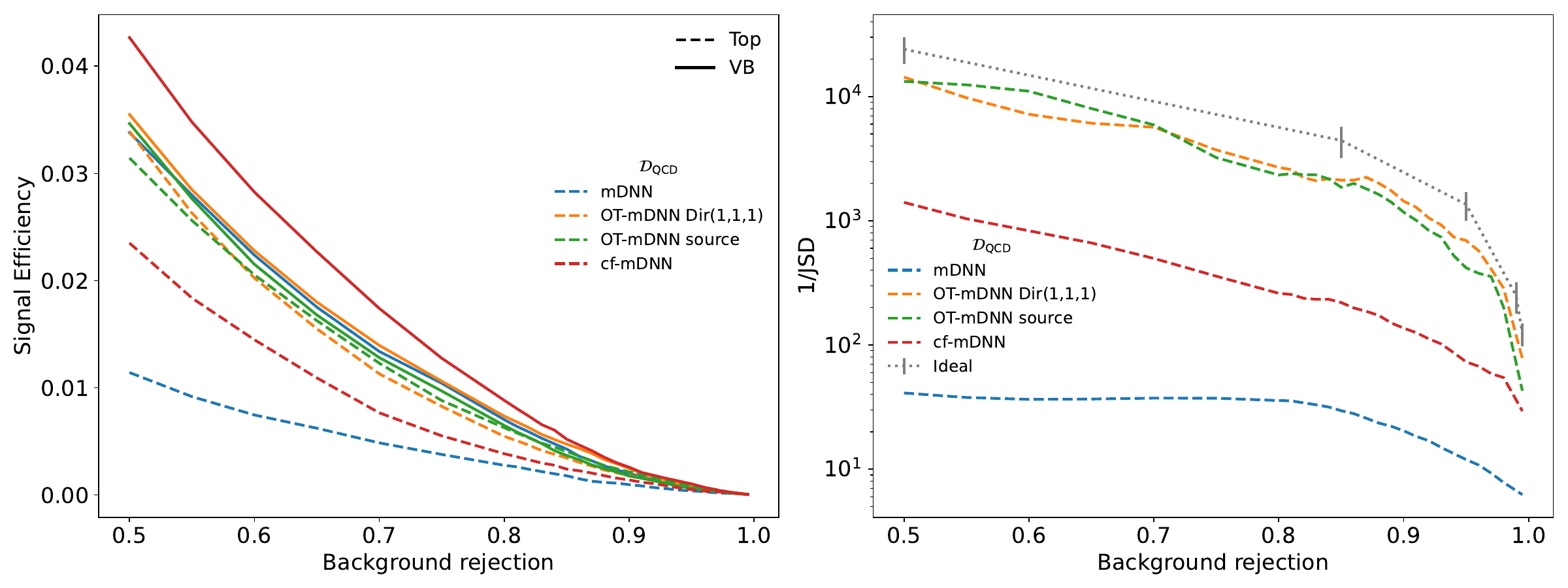}
        \caption{$\mathcal{D}_\mathrm{QCD}$ discriminant}
        \label{fig:qcd_3d_decorrelation_performance}
    \end{subfigure}%
    \\
    \begin{subfigure}[t]{1\textwidth}
        \centering
        \includegraphics[width=1\textwidth]{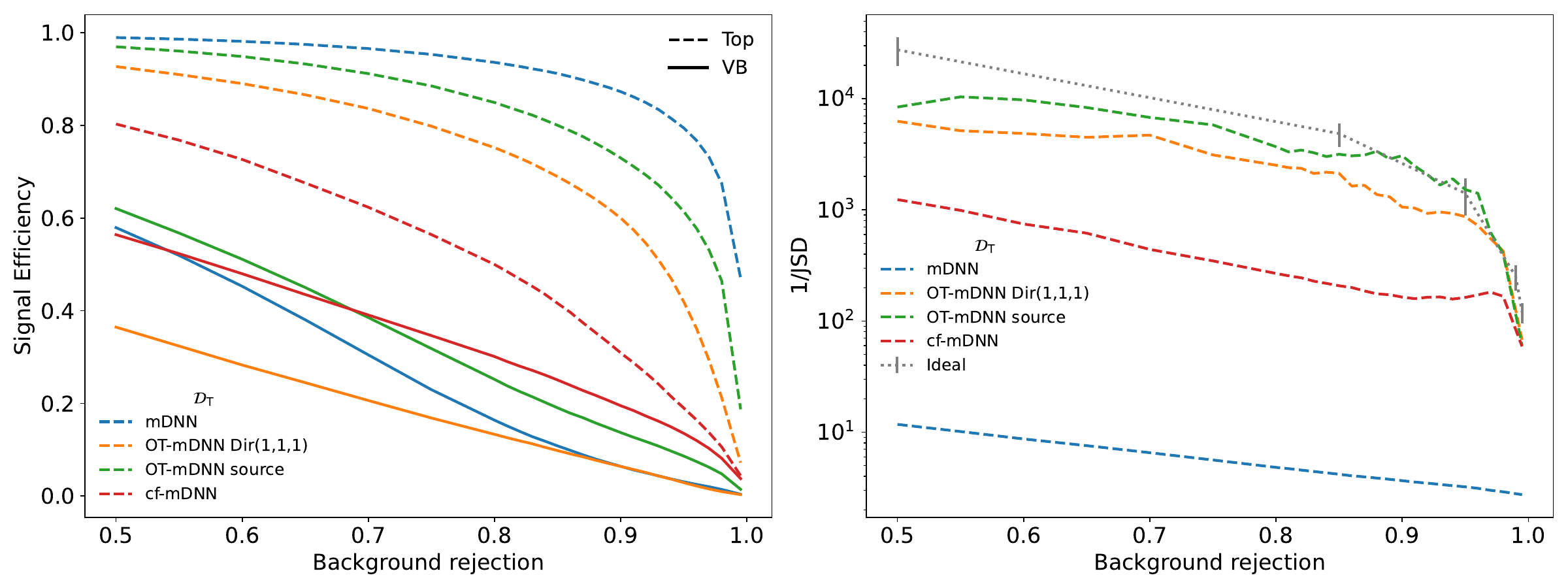}
        \caption{$\mathcal{D}_\mathrm{Top}$ discriminant}
        \label{fig:top_3d_decorrelation_performance}
    \end{subfigure}
    \\
    \begin{subfigure}[t]{1\textwidth}
        \centering
        \includegraphics[width=1\textwidth]{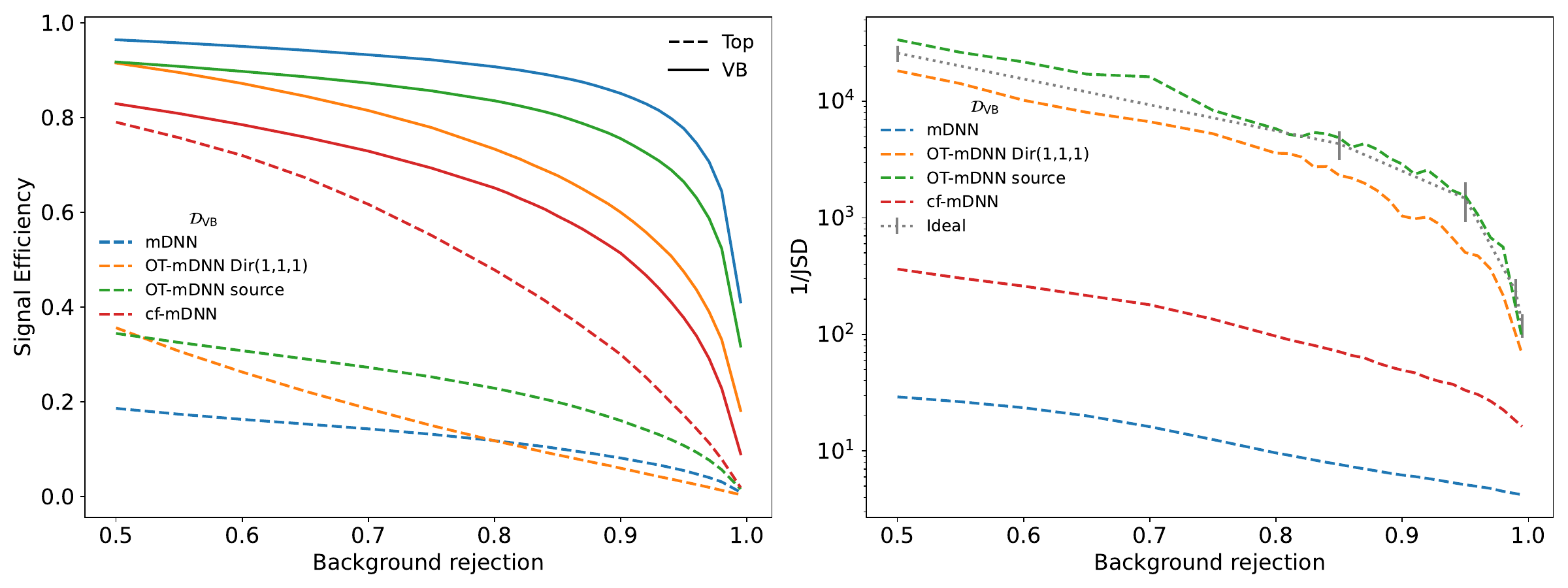}
        \caption{$\mathcal{D}_\mathrm{VB}$ discriminant}
        \label{fig:w_3d_decorrelation_performance}
    \end{subfigure}
    \caption{The signal efficiency and 1/JSD are measured at various levels of background rejections with the different discriminators from Eq.~\ref{eq:3d_discriminate}.
    The left column shows the signal efficiency. As we only consider Top and VB as signal, only the Top and VB signal efficiency are shown.
    The right column shows the measured background sculpting for different selections. The 1/JSD is measured only on QCD jets.}
    \label{fig:3d_decorrelation_performance}
\end{figure*}

To assess the decorrelation performance of the methods, we look at the background sculpting and signal efficiency of the discriminators after applying decorrelation.
The two methods are compared to the initial distributions in Fig.~\ref{fig:3d_decorrelation_performance}.
We will be testing Cnots applied to the classifier outputs (OT-mDNN) with two different base distributions. OT-mDNN Dir(1,1,1) uses 3D logit-Dirichlet as a base distribution with the concentration parameters set to one, the OT-mDNN source uses the original mDNN scores as a base distribution, and the cf-mDNN uses a normal distribution as a base distribution. All methods are trained in logit space and normalised with softmax during evaluation.
In Fig.~\ref{fig:3d_decorrelation_performance} we see both OT models outperform the cf-mDNN for all three discriminants, especially at high background rejection.
Here the decorrelated OT-mDNN scores follow the ideal case of no sculpting.
We also see that, in addition to the reduced levels of sculpting, the OT-mDNN models consistently have a higher signal efficiency than cf-mDNN for the discriminator optimised for the target jet type.

\begin{figure*}[htpb]
    \centering
    \begin{subfigure}[t]{0.32\textwidth}
        \centering
        \includegraphics[width=1\textwidth]{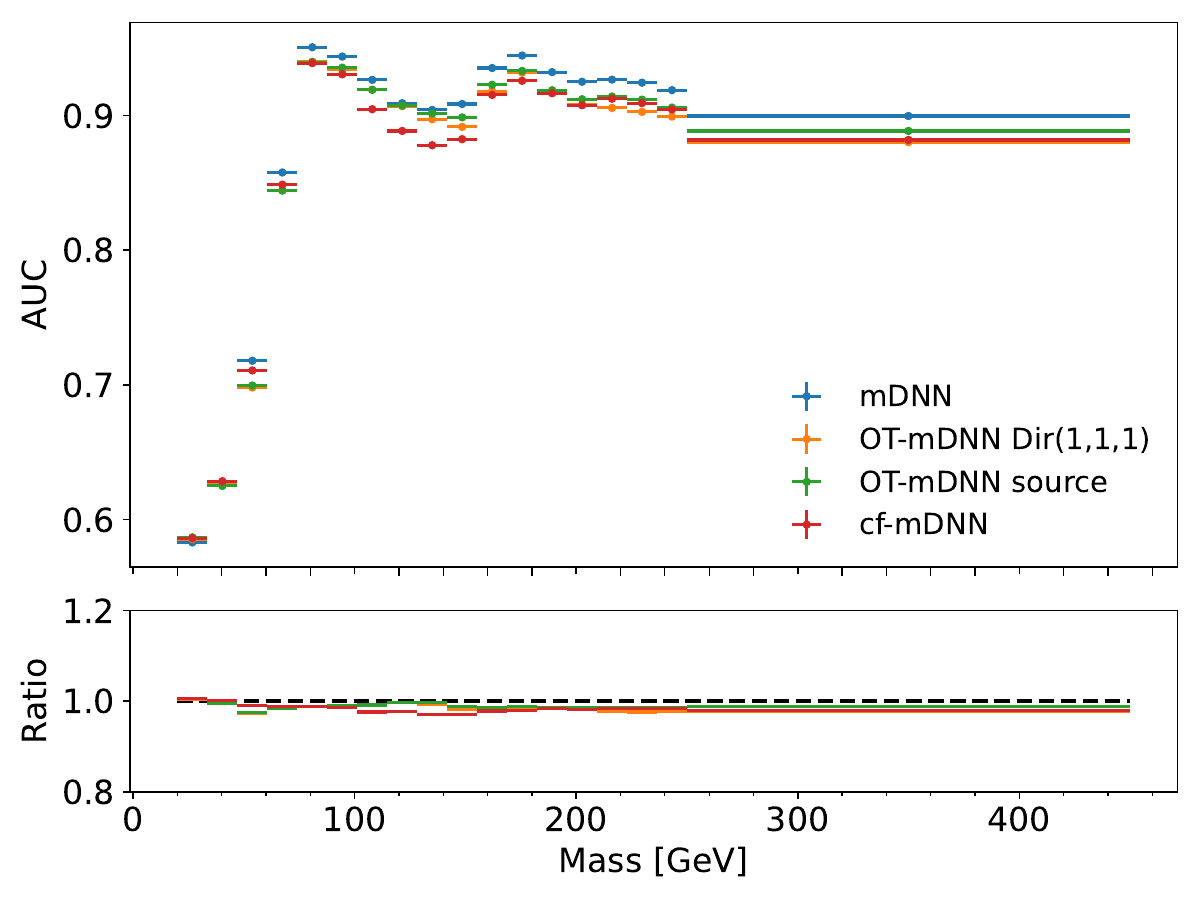}
    \end{subfigure}%
    \,
    \begin{subfigure}[t]{0.32\textwidth}
        \centering
        \includegraphics[width=1\textwidth]{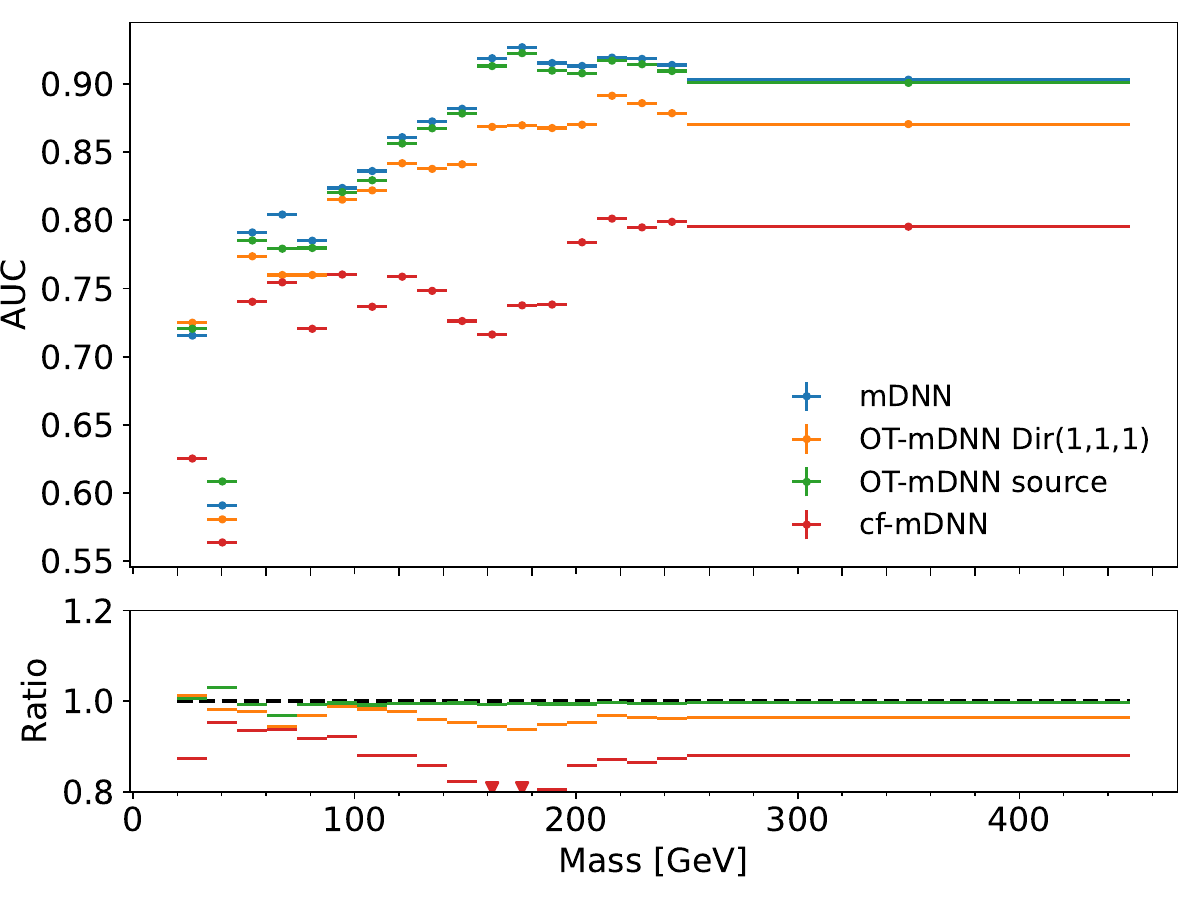}
    \end{subfigure}
    \,
    \begin{subfigure}[t]{0.32\textwidth}
        \centering
        \includegraphics[width=1\textwidth]{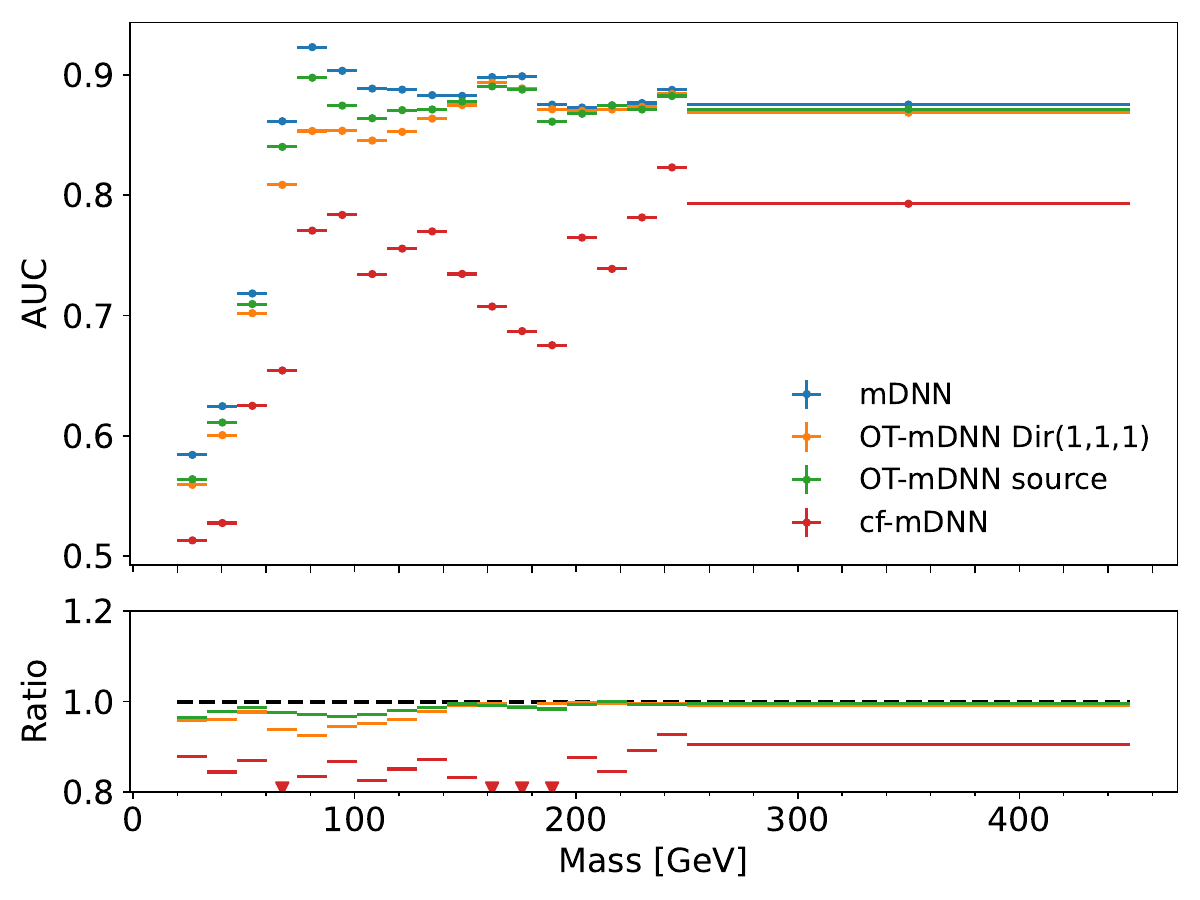}
    \end{subfigure}
    \caption{Comparison between the AUC before and after decorrelation as a function of mass for (a) $\mathcal{D}_\mathrm{QCD}$, (b) $\mathcal{D}_\mathrm{Top}$ and (c) $\mathcal{D}_\mathrm{VB}$.
    The ratio is calculated between the original AUC and the decorrelated ones.}
    \label{fig:3d_AUC}
\end{figure*}

The poor decorrelation performance of the cf-mDNN lies in its unconstrained transport maps, which do not contain the restriction that the map must be monotone and therefore allows order swapping.
However, this is not the case for the OT approach, which is order preserving by construction.
This is also clear from the differences in AUC as a function of jet mass shown in Fig~\ref{fig:3d_AUC}, where the integrated performance of cf-mDNN declines significantly compared to the original distribution. We also see that the performance for the two OT-mDNN methods is close to the optimal performance of the original distribution. The small discrepancy may be due to the finite binning size of the integral.

\subsubsection*{Order preservation}

The measure of order preservation is simple to quantify in 1D,
however, it becomes non-trivial in higher dimensions.
For an order preserving map, we expect to observe no rapid fluctuation in the transport map and minimum curl in the transport field.
We attempt to visualise the level of order preservation by depicting the gradient of the transport maps across the input space.
Low and smoothly changing gradients indicate a smooth and order preserving map.
However, large gradients or abrupt changes do not necessarily indicate where the order preservation is broken.

In order to compare the transport maps between OT-mDNN and cf-mDNN, a shared base distribution is required. We choose a normal distribution as the base distribution for both. 
In order to reintroduce unitarity per event, the outputs are rescaled using a softmax activation on the three outputs, ensuring two degrees of freedom. We then visualise the transport maps in 2D by using $p_{VB}$ and $p_{T}$. 


\begin{figure*}[htpb]
    \centering
    \includegraphics[width=1\textwidth]{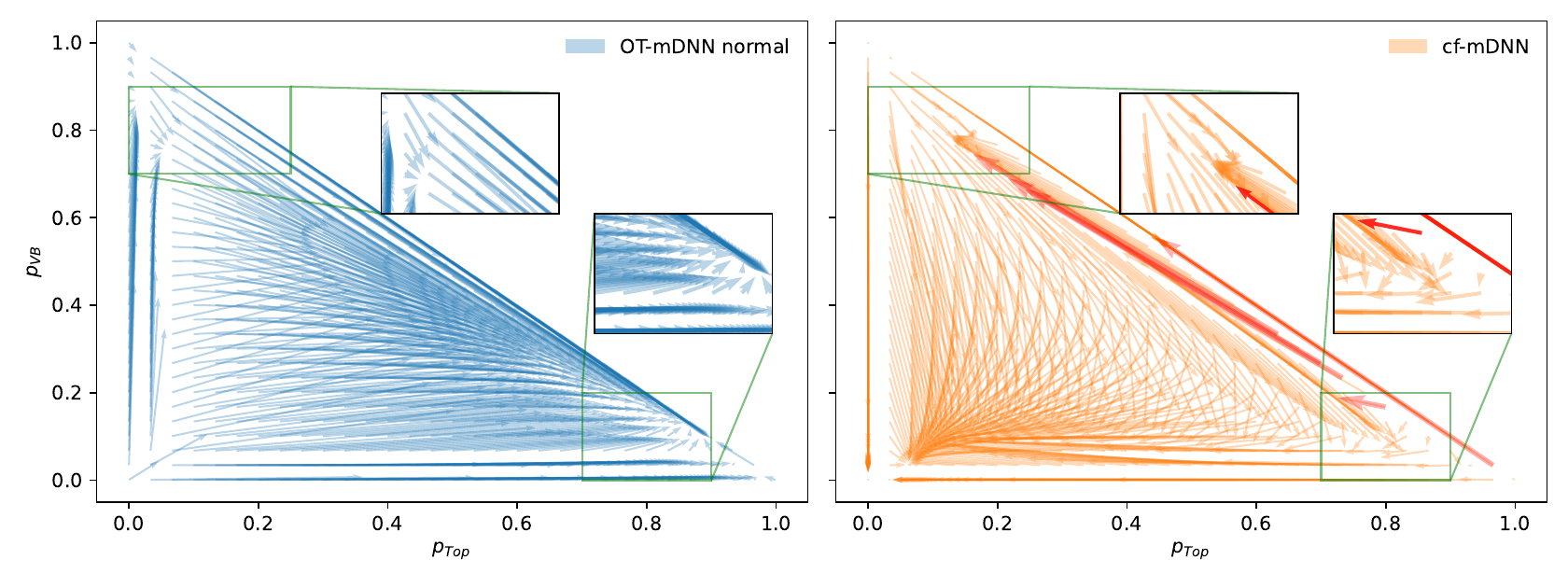}
    \caption{By taking the difference between points before and after decorrelation, we can calculate the displacement vector and construct the transport maps for both methods. These displacement vectors are visualised in this quiver plot. The transport maps are mass dependent, so to compare maps, the same mass value has to be chosen. For these two specific transport maps, the W-boson mass is chosen. The red arrows are the cf-mDNN displacement vectors that are above the maximum magnitude differences of the OT-mDNN seen in Fig.~\ref{fig:hist_magnitude_difference}.}
    \label{fig:displacements_vectors}
\end{figure*}
Taking the difference between correlated and decorrelated scores, we can calculate a displacement vector that indicates the direction and amount a given point is transported during decorrelation. As the scores are in Dirichlet space, one of the three dimensions becomes redundant and can be dropped, enabling us to show the transport maps in 2D.
The displacement vectors are shown in Fig.~\ref{fig:displacements_vectors} for OT-mDNN and cf-mDNN.
We show that in some regions the transport map of the cf-mDNN changes rapidly and overlaps into other regions, whereas the OT-mDNN has smoother transitions.
\begin{figure*}[htpb]
    \centering
    \includegraphics[width=0.7\textwidth]{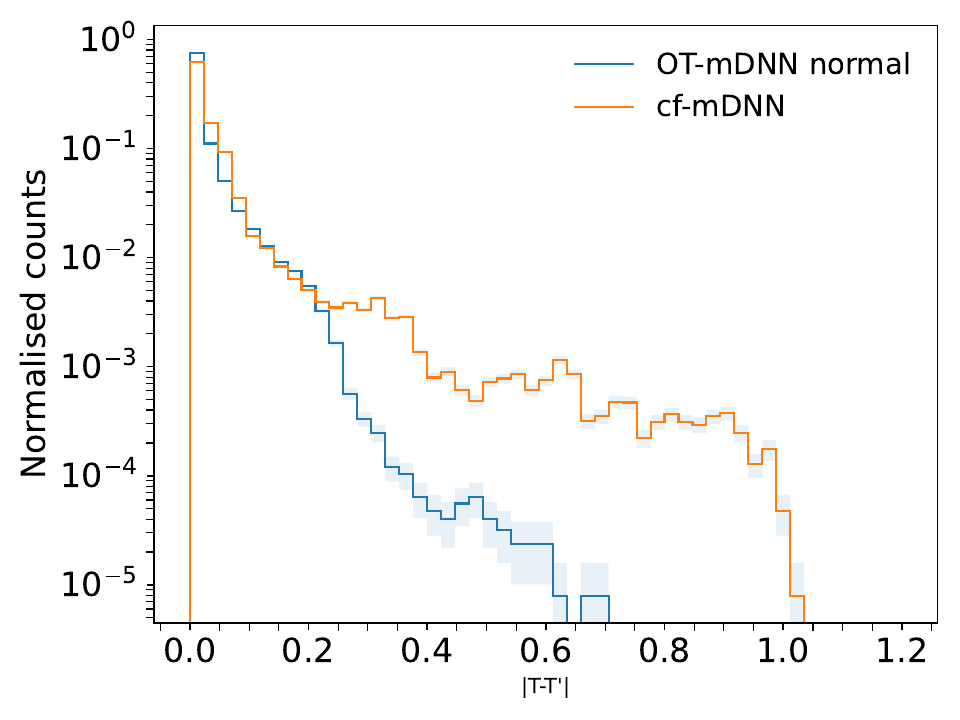}
    \caption{The magnitude of the difference between the original displacement vectors and their small deviation counterpart.}
    \label{fig:hist_magnitude_difference}
\end{figure*}
We will attempt to outline these rapid transitions by sampling random positional variations of $2\%$ deviation from the original points in Fig.~\ref{fig:displacements_vectors}. We then measure the distance between displacement vectors of the original point and a small deviation. A large difference indicates a rapid change in the transport map, which can result in non-order preserving transport.
In Fig.~\ref{fig:hist_magnitude_difference}, the histogram of magnitudes is shown. 
Here, $1 \%$ of the cf-mDNN displacement vectors have a magnitude larger than the maximum magnitude of the OT-mDNN. These displacement vectors are indicated in red in Fig.~\ref{fig:displacements_vectors}. 
These red arrows strongly indicate that we have regions in the cf-mDNN transport map where the monotonicity is broken, which results in the low signal efficiency we saw in Fig.~\ref{fig:3d_decorrelation_performance}. 

\begin{figure*}[htpb]
    \centering
    \includegraphics[width=1\textwidth]{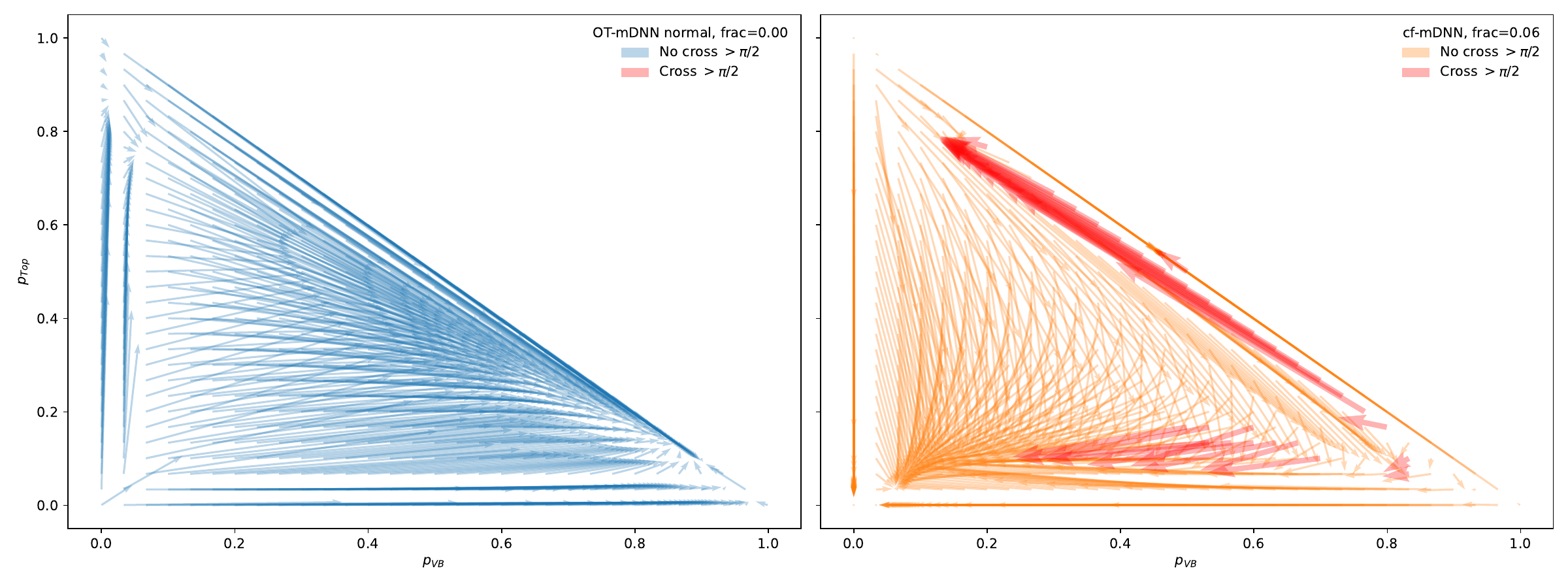}
    \caption{The displacement vectors for the transport maps of the OT-mDNN (left) and cflow (right). An invariant mass of 80.377 [GeV] is selected for these transport maps. Displacement vectors that cross each other by an angle greater than $\pi/2$ are highlighted in red, highlighting where the transport is non-convex. }
    \label{fig:angle_diff}
\end{figure*}


As another measure of order preservation, we try to quantify areas of the transport map where order swapping may occur.
In the case of a perfectly monotonic transport in two dimensions, the transport field should be convex, we postulate that for this to hold no transport paths should cross with an angle greater than $\pi/2$.
In Fig.~\ref{fig:angle_diff}, we show the same transport maps as in Fig.~\ref{fig:displacements_vectors}, but highlight transport vectors that cross any other transport vector by an angle greater than $\pi/2$.
Here we see that this does not hold for cflows, however, it does hold for OT-mDNN.

These rapid changes in the transport map will break the monotonicity. However, a clear quantification of monotonicity in higher dimensions are not known to us.

%% file: includes/conclusion.tex
\section{Conclusion}
In this work we have introduced a novel method for decorrelating feature spaces correlated to protected attributes by finding the optimal transport map between the correlated space and a decorrelated space.
This map is cosntructed as the gradient of a partially input convex neural network, ensuring it is monotonic by construction. 
We study the decorrelation performance of our approach in comparison to the state-of-the-art for jet tagging at the LHC.
Conditional normalising flows~\cite{sam} have demonstrated success in decorrelating 1D distributions, with our approach reaching similar levels of performance.
However, Cnots achieves state-of-the-art performance outperforming the normalising flows at decorrelating higher dimensional feature spaces.
This increase in performance is achieved due to the enforced monotonicity in the architecture, which, although present in 1D for the normalising flows, is not enforced in higher dimensions.
Furthermore, Cnots can perform decorrelation with an arbitrary distribution chosen as the target of the transport.


The application of decorrelation is not restricted to classifier outputs, and due to the state-of-the-art performance in decorrelating higher dimensional feature spaces, Cnots should result in improved performance for tasks which require decorrelation of input feature spaces.




The code is publicly available at \url{https://github.com/malteal/ot-decorrelation}.

%% file: includes/acknowledgement.tex
\section*{Acknowledgements}
The authors would like to acknowledge funding through the SNSF Sinergia grant called Robust Deep Density Models for High-Energy Particle Physics and Solar Flare Analysis (RODEM) with funding number CRSII$5\_193716$ and the SNSF project grant 200020\_212127 called "At the two upgrade frontiers: machine learning and the ITk Pixel detector". We would also like to thank Matthew Leigh for training the classifier used in the analysis and Samuel Klein and Chris Pollard for useful discussions and feedback.

%% file: includes/appendix.tex
\appendix
\section*{Appendix}
\addcontentsline{toc}{section}{Appendices}
\renewcommand{\thesubsection}{\Alph{subsection}}
\subsection{Multiclass classifier} \label{sec:mDNN}
The transformer use 3 self-attention layers and 2 cross-attention layers, each with an embedding dimension of 128 and 16 attention heads. 
The 7 node features and 5 edge features are embedded into this space each using a dense network with a single hidden layer of 256 neurons.
We also use the same dense network shape for the residual updates inside the transformer, and to extract the final 3-value discriminant from the cross attention blocks. 
Conditioning on the 2 high level variables is achived by concatenating them to the input of all dense networks use in this setup. 
We use the LeakyReLU activation and a dropout probability of 0.1 in all linear layers.
We trained the transformer for 10 epochs using the AdamW optimiser with a learning rate of 0.001 and a weight decay strength of 0.0001.

\subsection{Cflow} \label{sec:cflows}
The cflow for binary decorrelation is training using Adam optimiser for 300 epochs with initial learning rate at $0.0005$ that annealed to zero using cosine annealing with batch size of $512$. The cflows is constructed by 5 rational quadratic spline layers with 12 bins in each layer using a uniform[0,1] or a normal distribution as a base distribution for either binary and multiclass decorrelation, respectively. The invariant mass is transformed to logarithmic mass.
The multiclass decorrelation cflow is trained with same training setup as the binary one. However it is constructed by 6 rational quadratic spline layers with 12 bins bounded within [-3.5, 3.5] in each layer using a normal distribution as a base distribution. Outside the bins, the map is linear. 
Everything else followed the default settings from Ref.~\cite{nflows}.
\subsection{PICNN} \label{sec:PICNN}
The architecture for the PICNNs use in both binary and multiclass decorrelation can be seen in Tab.~\ref{tab:PICNN}. We always train our transport in logit space with logarithmic invariant mass as conditions. For both multiclass and binary decorrelation we test the same two base distributions, a uniform[0,1] and the original source distribution.
\begin{table}[htpb]
    \begin{tabular}{lll}
    Category                  & Hyperparameter                & Value                          \\ \hline
    \multirow{9}{*}{Training} & Optimiser                     & AdamW(0, 0.9)                  \\
                              & Epochs                        & 150                            \\
                              & F per g                       & 4                              \\
                              & G per f                       & 10                             \\
                              & Graident clipping             & 5                              \\
                              & Number of batches per epoch   & 512                            \\
                              & Batch size   & 1024                            \\
                              & Learning rate                 & 0.001                          \\
                              & Scheduler                     & CosineAnnealingLR(0.001, 5e-5) \\ \hline
    \multirow{5}{*}{Network}  & Convex layer size             & 64                             \\
                              & Non-convex layer size         & 8                              \\
                              & Number of layers              & 4                              \\
                              & Convex activation function    & Zeroed softplus                       \\
                              & Nonconvex activation function & Elu                            \\ \hline
    \end{tabular}
    \caption{The table show the hyperparameter for the $f$ and $g$ network use for our decorrelation method. Zeroed softplus is $SP(x)-SP(0)$}
    \label{tab:PICNN}
\end{table}



\subsection{Binary decorrelation} \label{sec:app_binary_decorre}
\begin{figure*}[htpb]
    \centering
    \begin{subfigure}[t]{0.45\textwidth}
        \centering
        \includegraphics[width=1\textwidth]{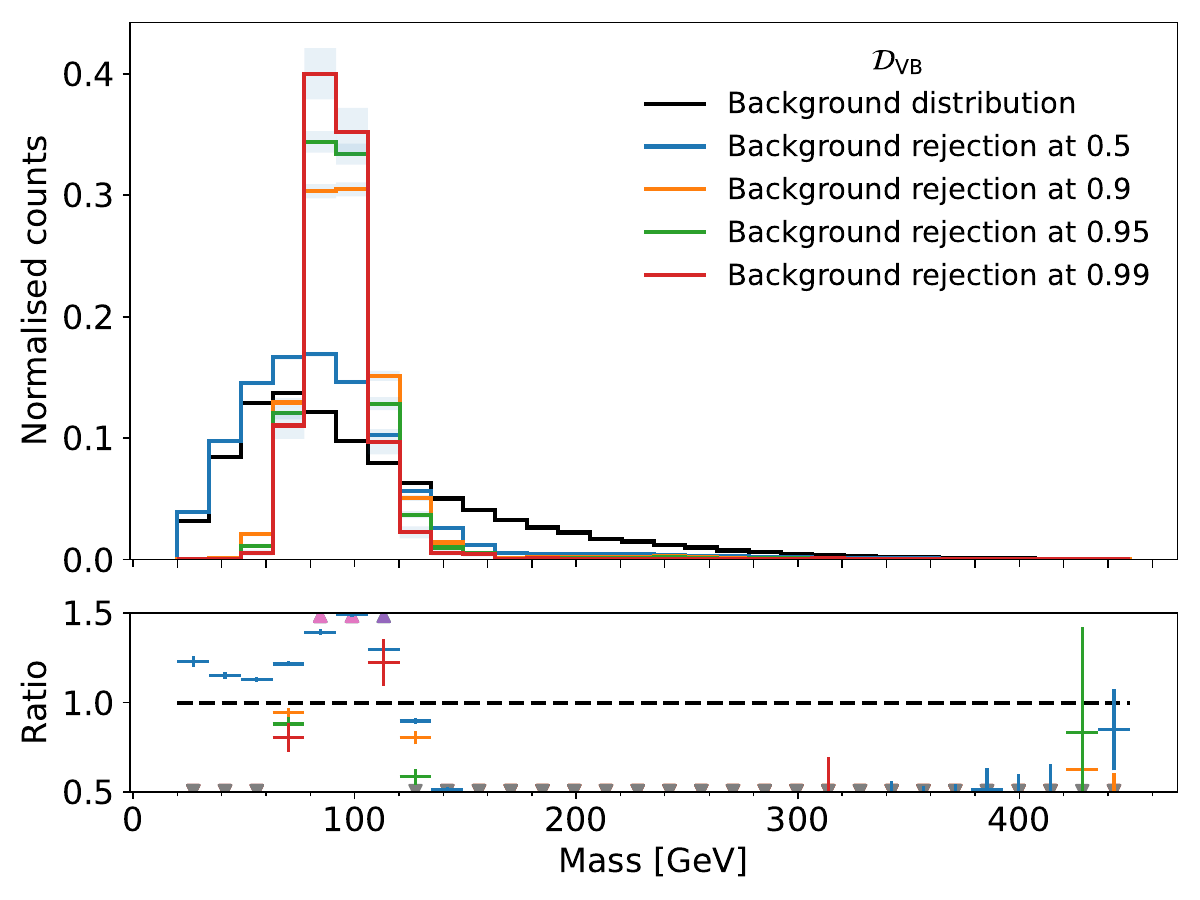}
        \caption{$\mathcal{D}_\mathrm{VB}$ mass sculpting}
    \end{subfigure}%
    \begin{subfigure}[t]{0.45\textwidth}
        \centering
        \includegraphics[width=1\textwidth]{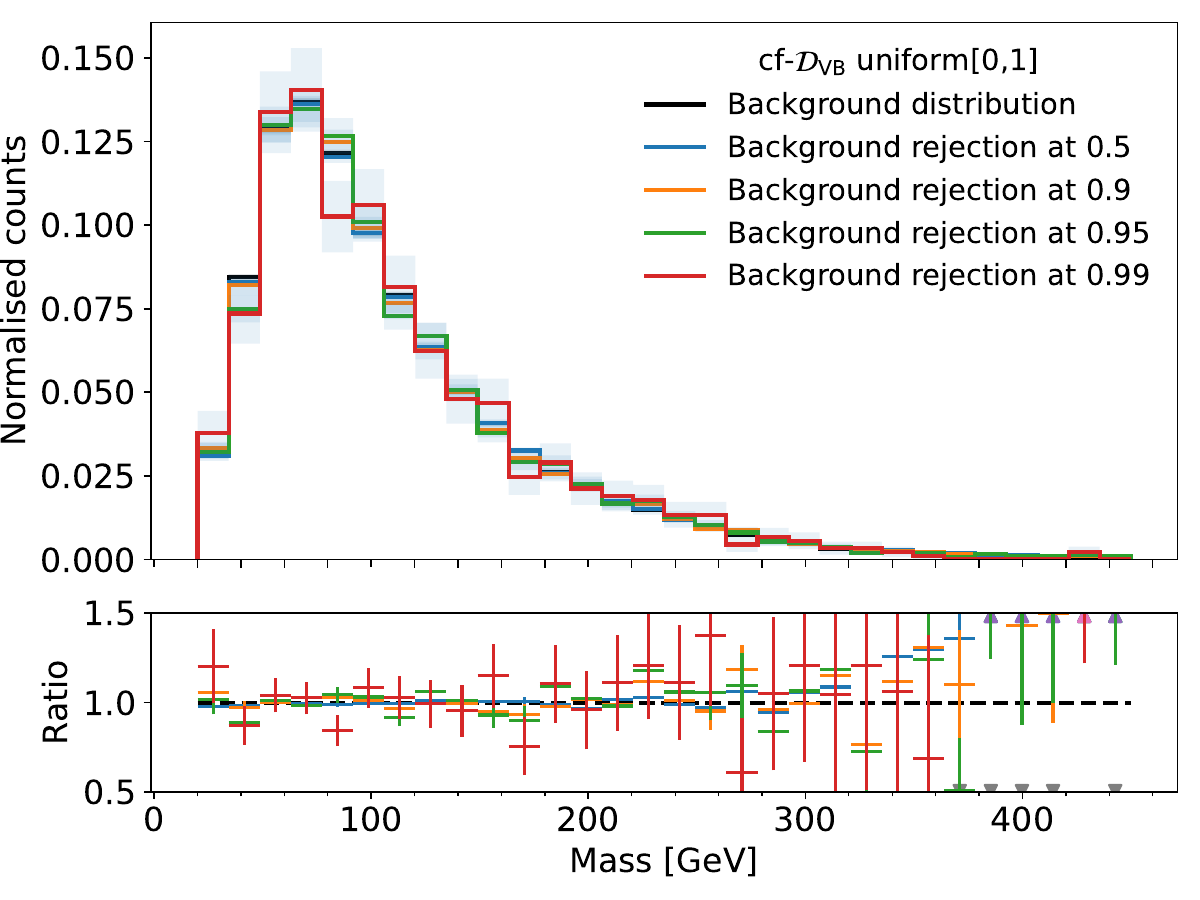}
        \caption{cf-$\mathcal{D}_\mathrm{VB}$ mass sculpting}
    \end{subfigure}
    \\
    \begin{subfigure}[t]{0.45\textwidth}
        \centering
        \includegraphics[width=1\textwidth]{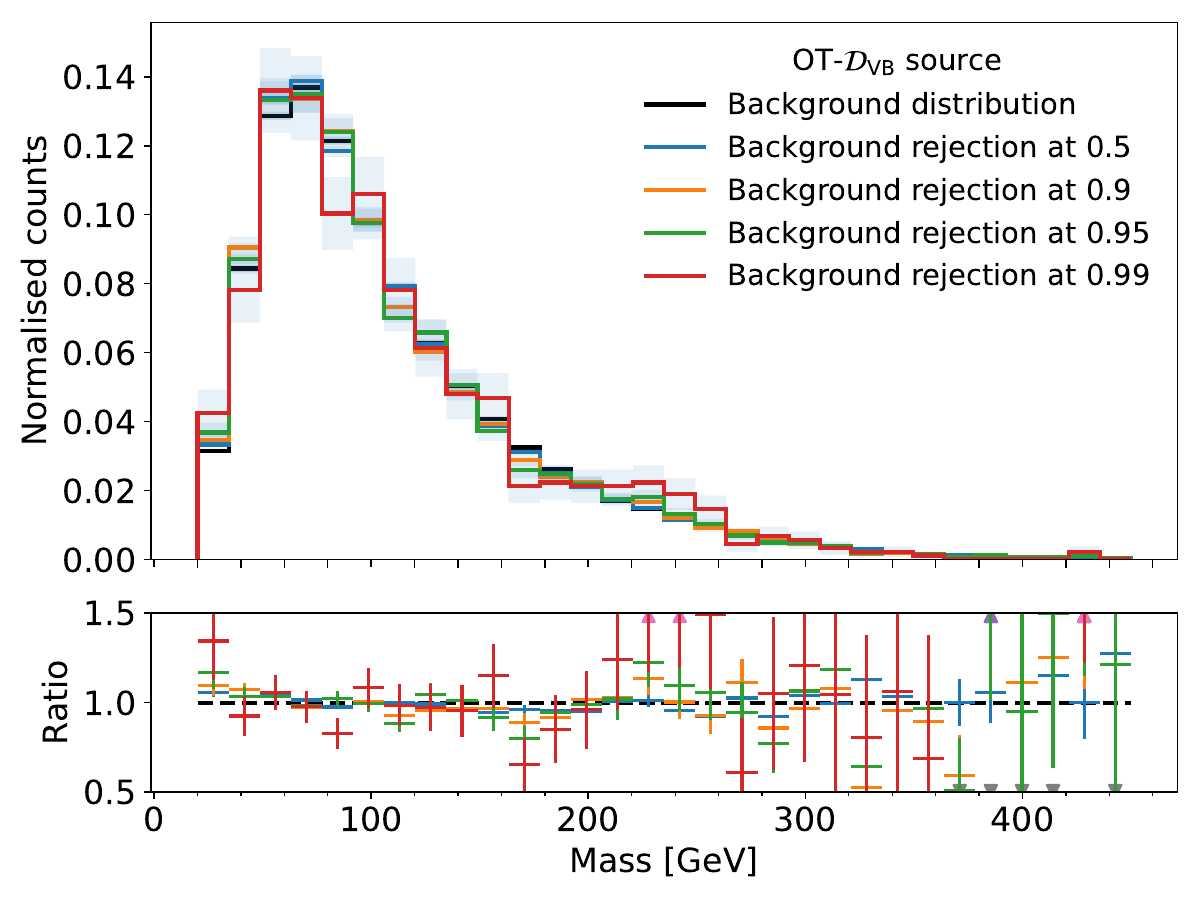}
        \caption{OT-$\mathcal{D}_\mathrm{VB}$ source mass sculpting}
    \end{subfigure}
    \begin{subfigure}[t]{0.45\textwidth}
        \centering
        \includegraphics[width=1\textwidth]{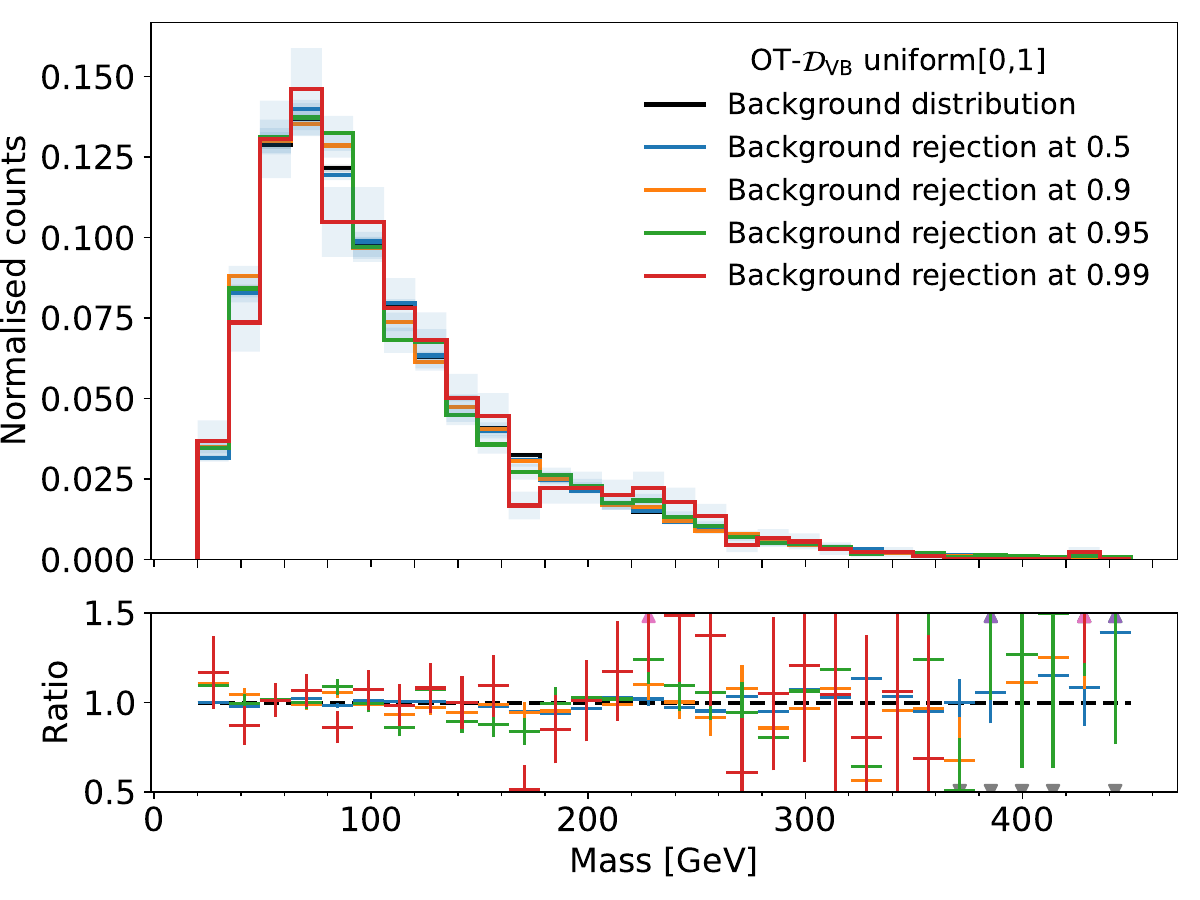}
        \caption{OT-$\mathcal{D}_\mathrm{VB}$ uniform mass sculpting}
    \end{subfigure}
    \caption{The figures show the mass sculpting given a background rejection for $\mathcal{D}_\mathrm{VB}$, OT-$\mathcal{D}_\mathrm{VB}$ and cf-$\mathcal{D}_\mathrm{VB}$.}
    \label[type]{fig:mass_sculpting}
\end{figure*}

To compare the 1/JSD values for the different methods, the binning and ranges between distribution has to be the same. We have chosen 50 bins equally spaced between 20 and 450 GeV for all mass sculpting figures. 
In Fig.~\ref{fig:mass_sculpting}, we see the mass sculpting at different background rejections. These are the distribution that are used to measure the 1/JSD values. We see that after decorrelating the scores of $\mathcal{D}_\mathrm{VB}$, the sculpting becomes non existent.

\subsection{Multiclass decorrelation} \label{sec:app_multi_decorre}

\begin{figure*}[htpb]
    \centering
    \begin{subfigure}[t]{0.33\textwidth}
        \centering
        \includegraphics[width=1\textwidth]{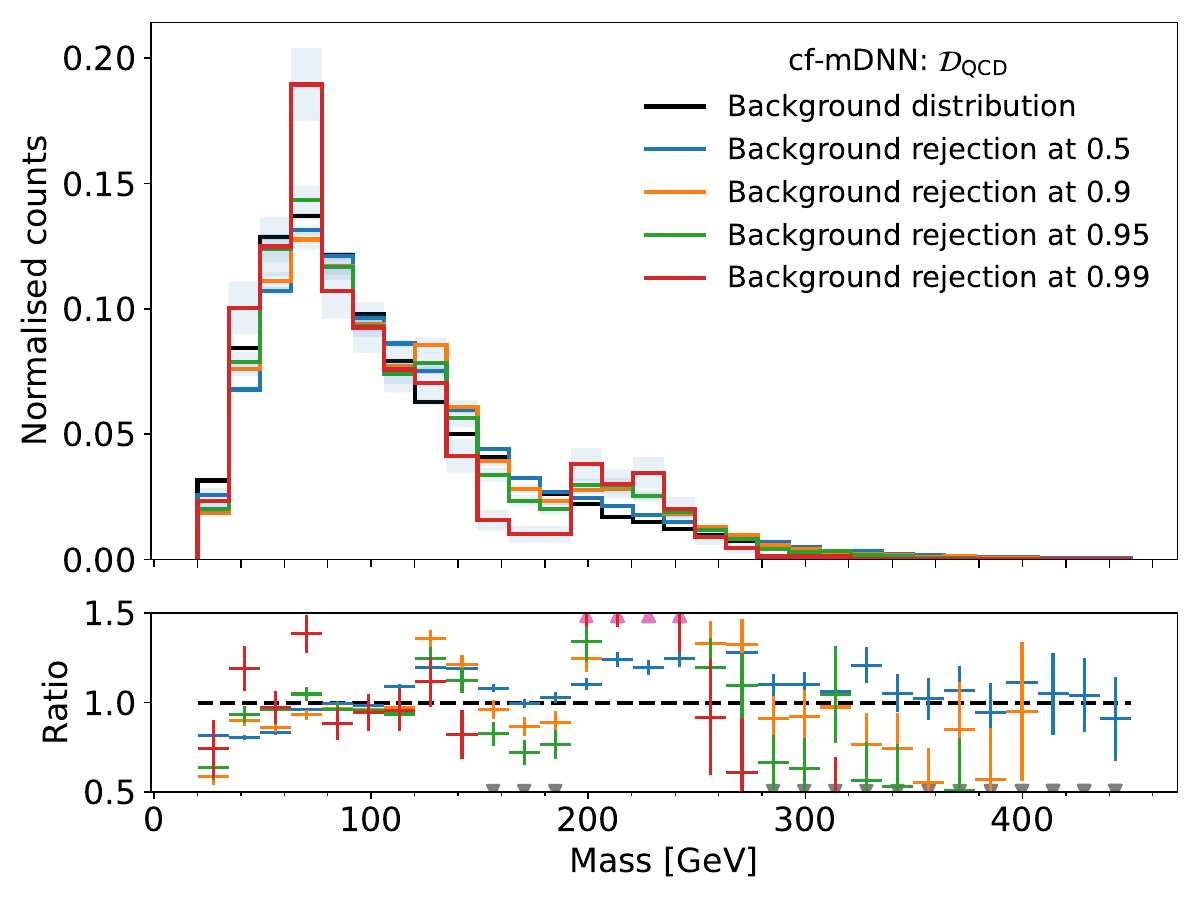}
        \caption{$\mathcal{D}_\mathrm{QCD}$ projection}
    \end{subfigure}%
    \begin{subfigure}[t]{0.33\textwidth}
        \centering
        \includegraphics[width=1\textwidth]{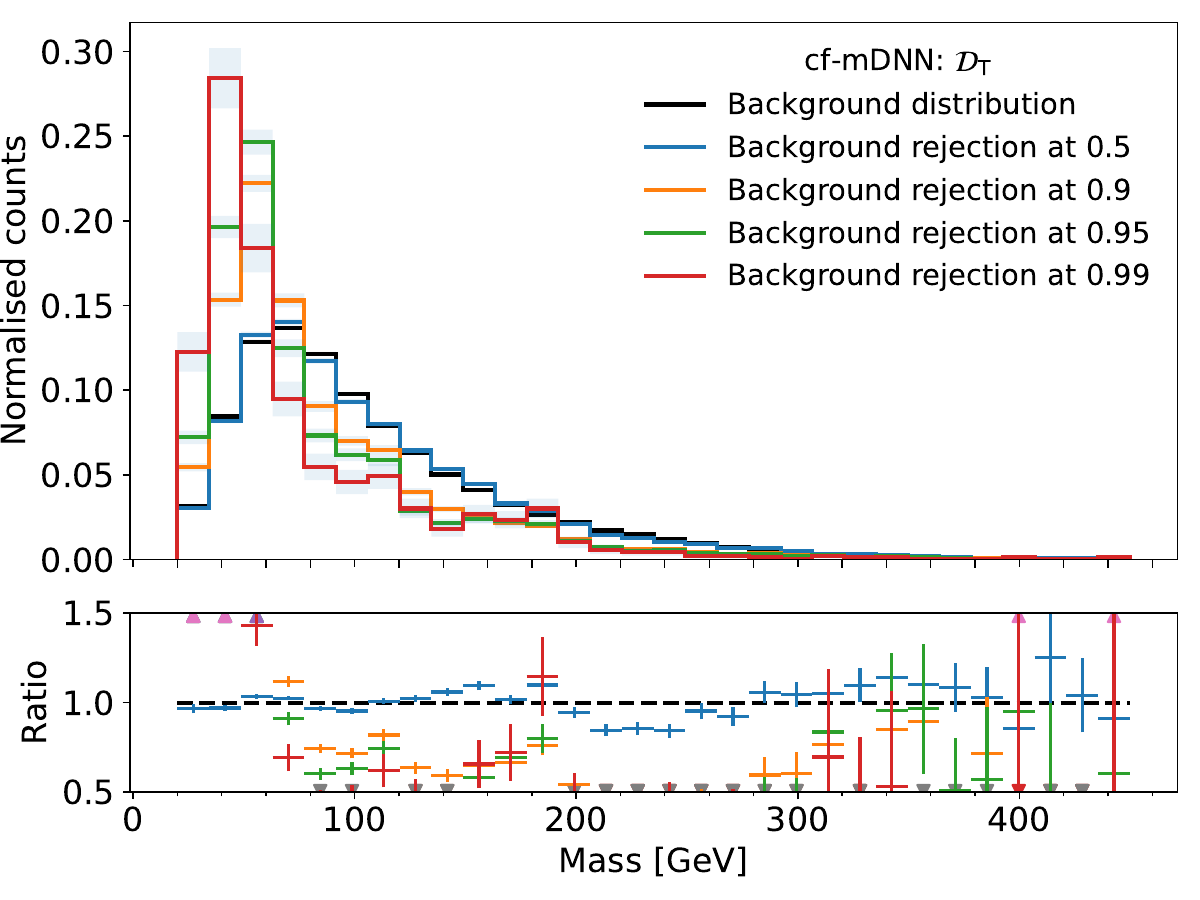}
        \caption{$\mathcal{D}_\mathrm{Top}$ projection}
    \end{subfigure}
    \begin{subfigure}[t]{0.33\textwidth}
        \centering
        \includegraphics[width=1\textwidth]{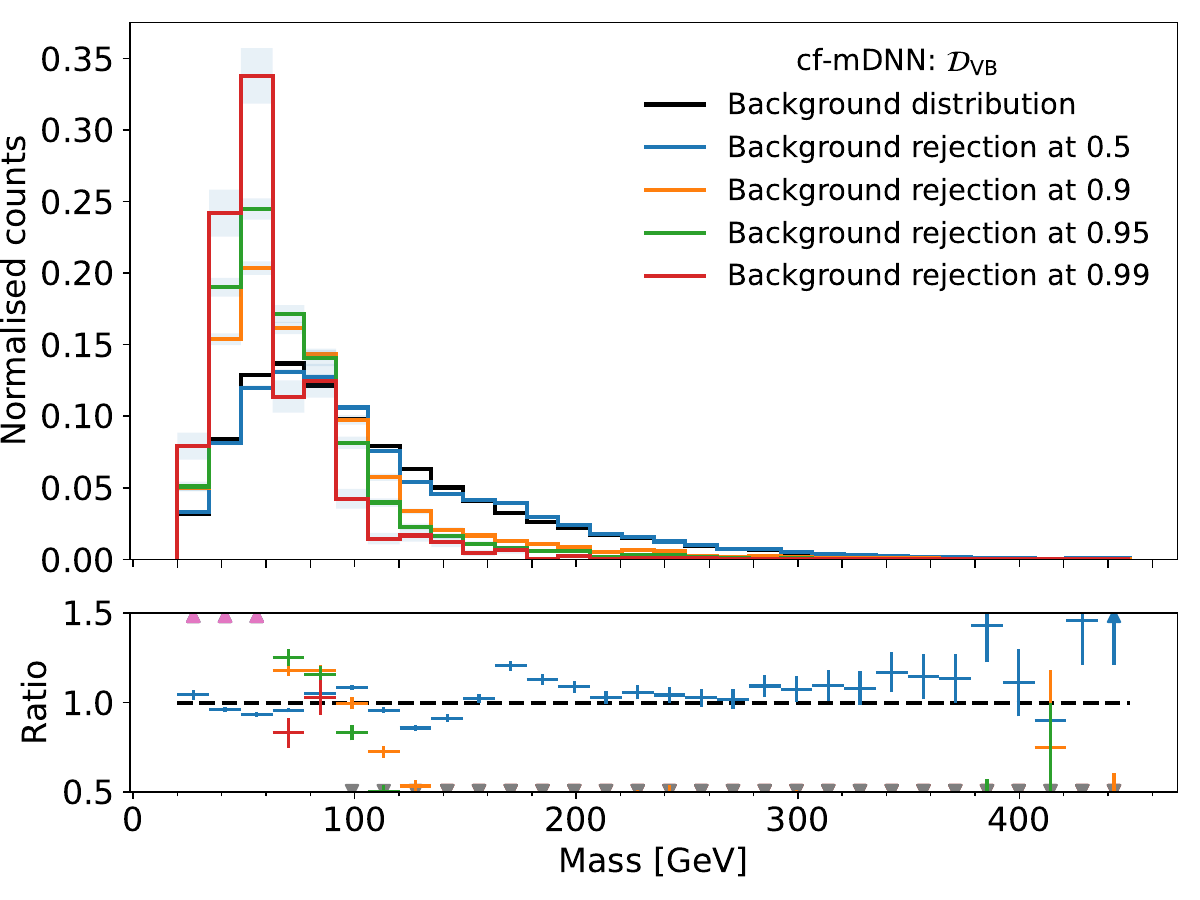}
        \caption{$\mathcal{D}_\mathrm{VB}$ projection}
    \end{subfigure}
    \caption{Mass distribution at background rejections of 50\%, 90\%, 95\% and 99\% after decorrelating the mDNN scores using the cflow method.}
    \label{fig:3d_original_mass_sculpting_cf}
\end{figure*}

\begin{figure*}[htpb]
    \centering
    \begin{subfigure}[t]{0.33\textwidth}
        \centering
        \includegraphics[width=1\textwidth]{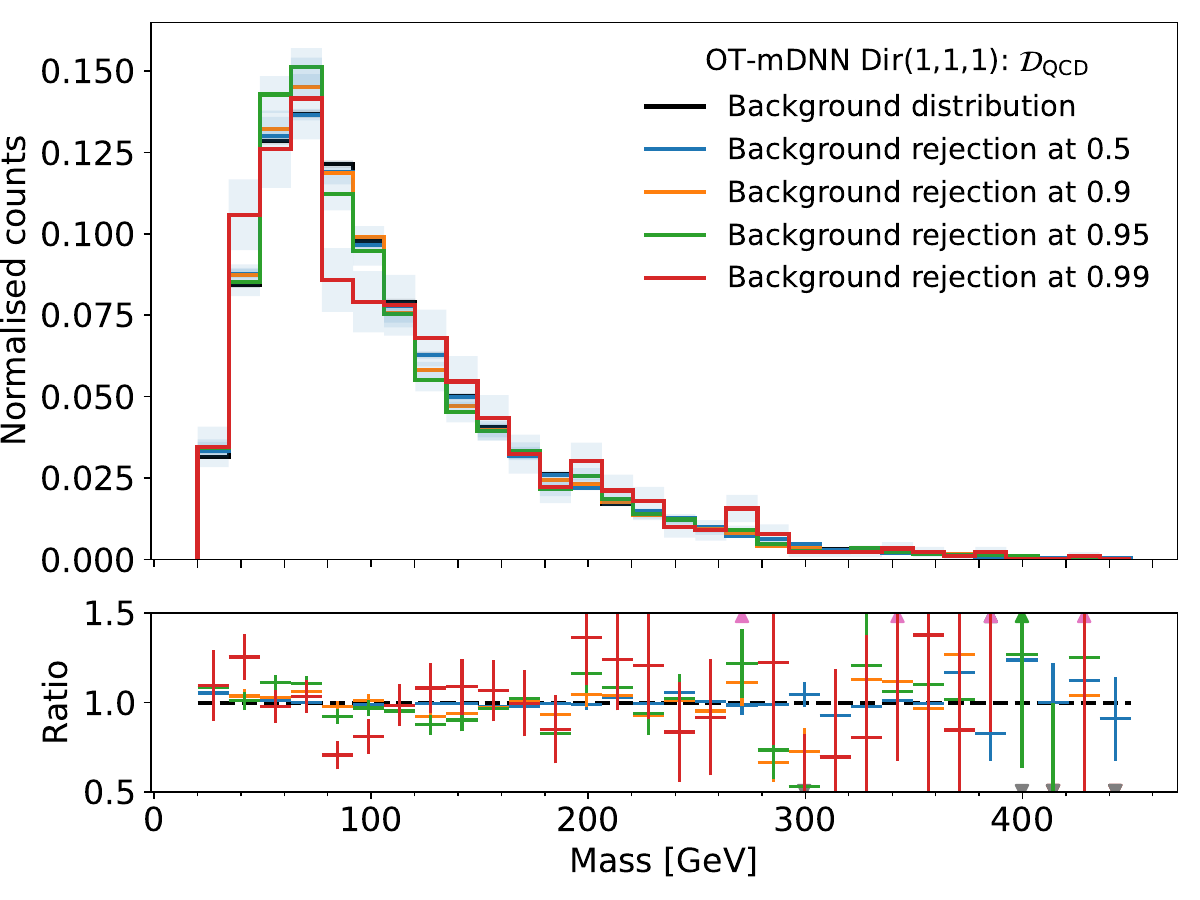}
        \caption{$\mathcal{D}_\mathrm{QCD}$ projection}
    \end{subfigure}%
    \begin{subfigure}[t]{0.33\textwidth}
        \centering
        \includegraphics[width=1\textwidth]{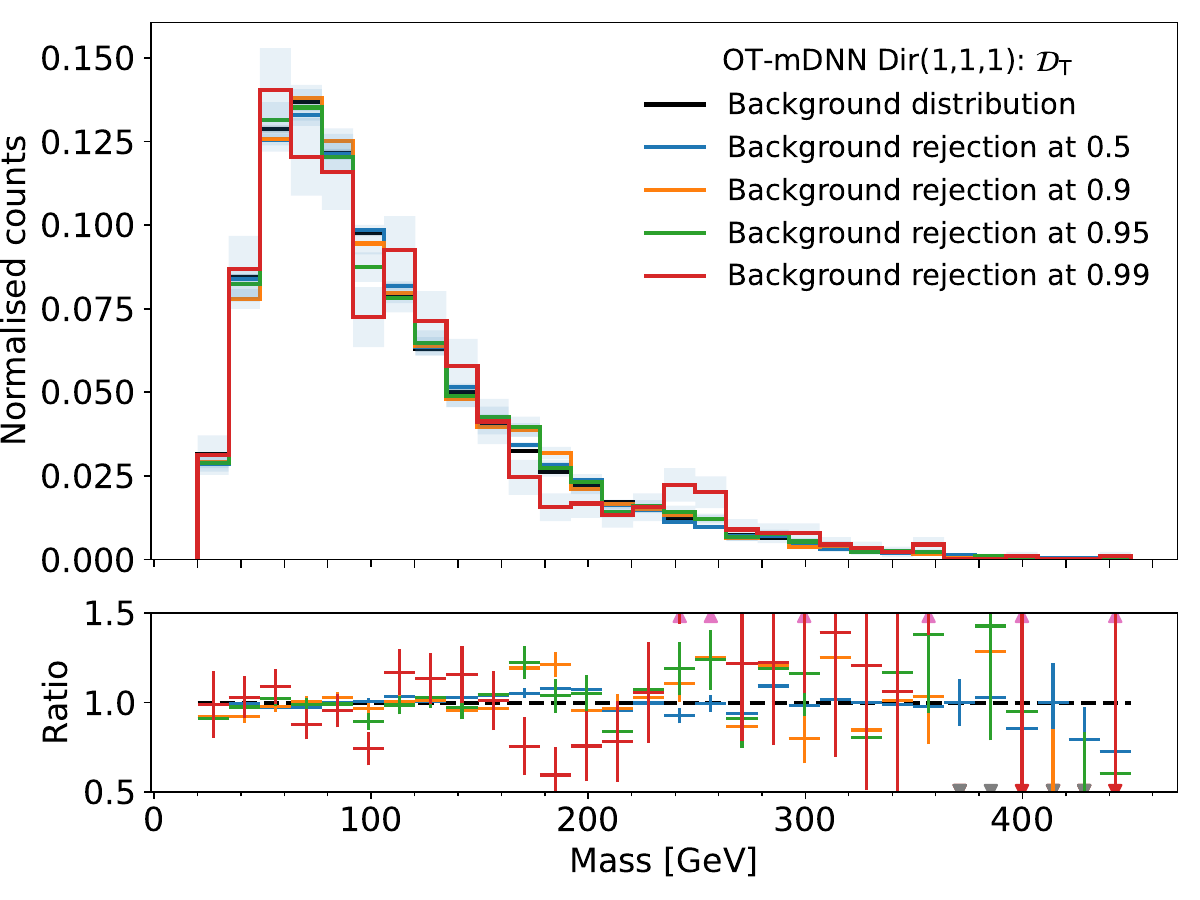}
        \caption{$\mathcal{D}_\mathrm{Top}$ projection}
    \end{subfigure}
    \begin{subfigure}[t]{0.33\textwidth}
        \centering
        \includegraphics[width=1\textwidth]{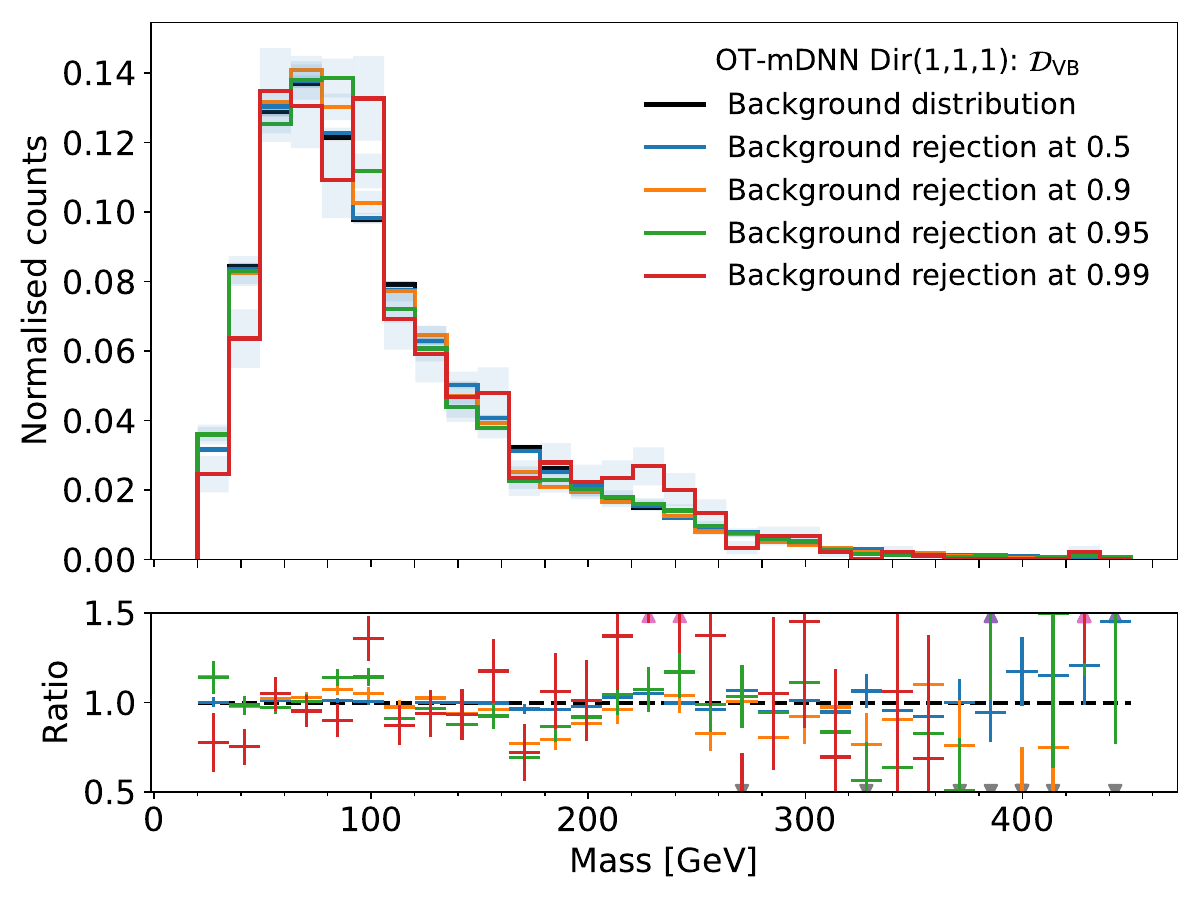}
        \caption{$\mathcal{D}_\mathrm{VB}$ projection}
    \end{subfigure}
    \caption{Mass distribution at background rejections of 50\%, 90\%, 95\% and 99\% after decorrelating the mDNN scores using the OT method with logit-Dirichlet as a base distribution.}
    \label{fig:3d_original_mass_sculpting_OT_dirichlet}
\end{figure*}

\begin{figure*}[htpb]
    \centering
    \begin{subfigure}[t]{0.33\textwidth}
        \centering
        \includegraphics[width=1\textwidth]{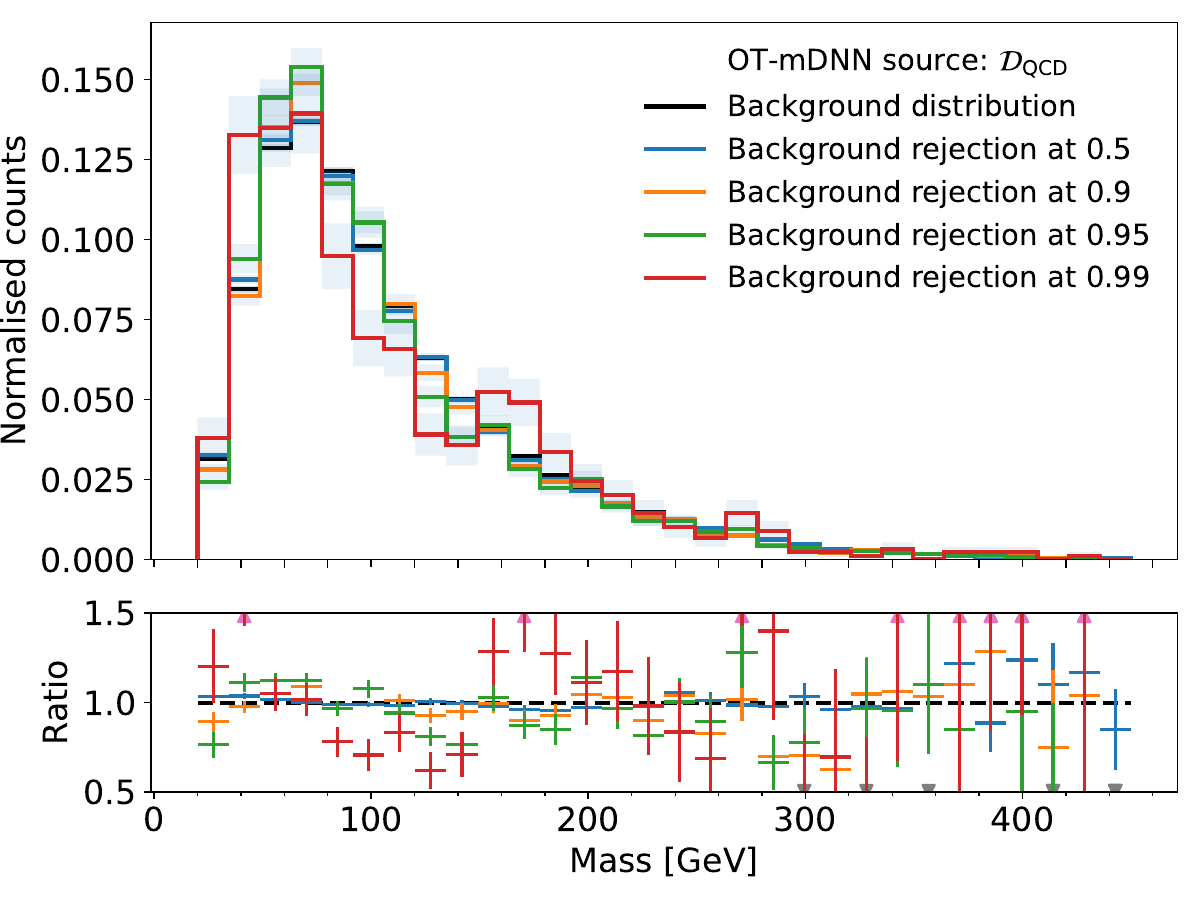}
        \caption{$\mathcal{D}_\mathrm{QCD}$ projection}
    \end{subfigure}%
    \begin{subfigure}[t]{0.33\textwidth}
        \centering
        \includegraphics[width=1\textwidth]{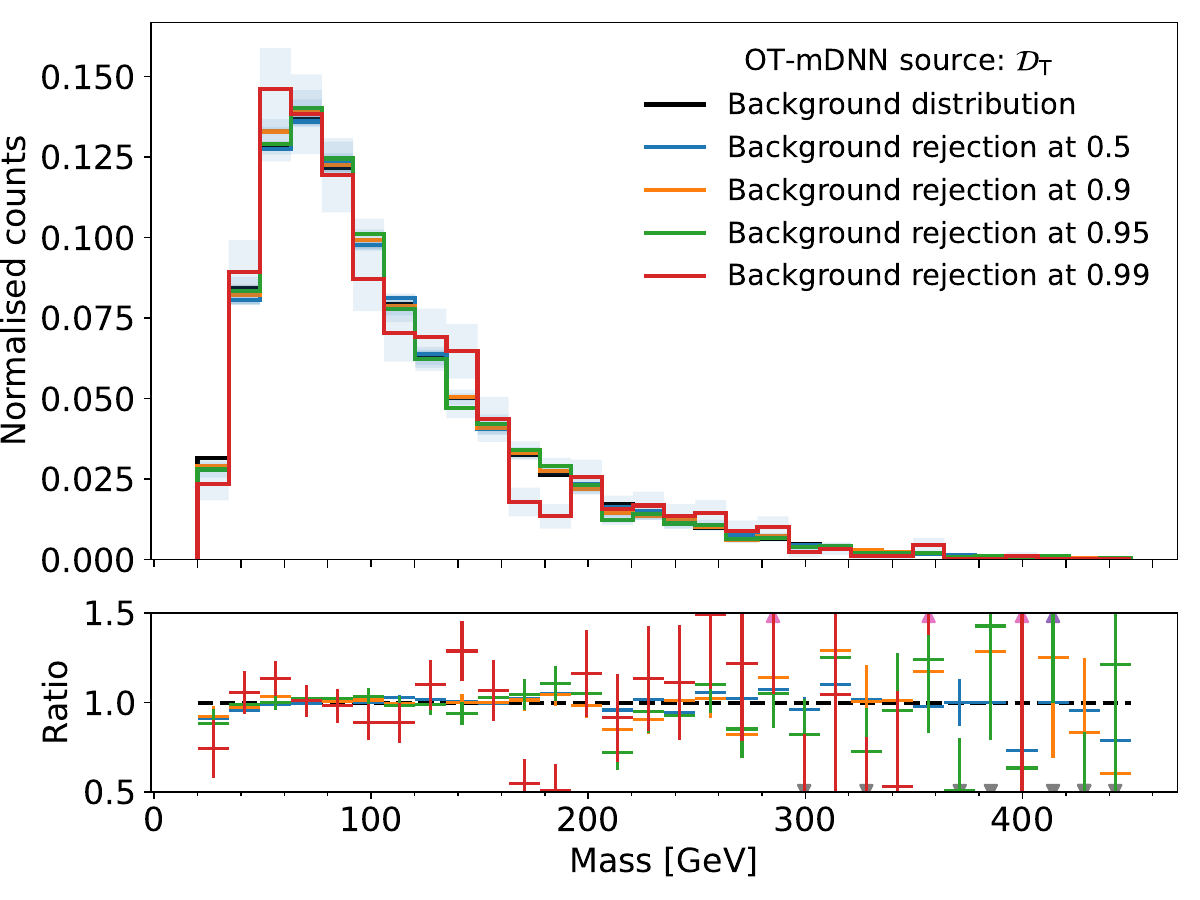}
        \caption{$\mathcal{D}_\mathrm{Top}$ projection}
    \end{subfigure}
    \begin{subfigure}[t]{0.33\textwidth}
        \centering
        \includegraphics[width=1\textwidth]{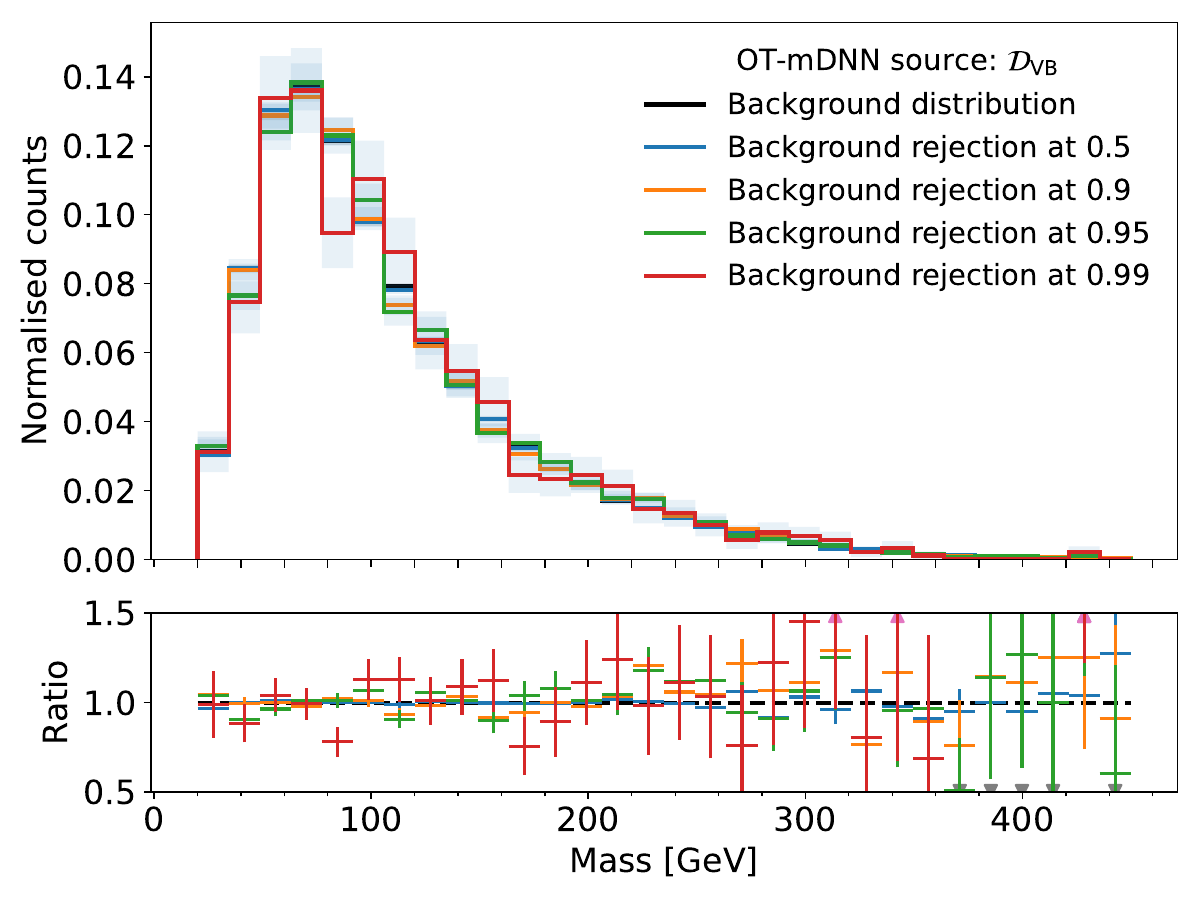}
        \caption{$\mathcal{D}_\mathrm{VB}$ projection}
    \end{subfigure}
    \caption{Mass distribution at background rejections of 50\%, 90\%, 95\% and 99\% after decorrelating the mDNN scores using the OT method with the original source distribution as a base distribution.}
    \label{fig:3d_original_mass_sculpting_OT_soure}
\end{figure*}

In Fig.~\ref{fig:3d_original_mass_sculpting}, we see the original mass sculpting for all three discriminators. In Fig.~\ref{fig:3d_original_mass_sculpting_cf},\ref{fig:3d_original_mass_sculpting_OT_dirichlet} and \ref{fig:3d_original_mass_sculpting_OT_soure}, we see the mass sculpting after using the decorrelation methods.

\begin{figure*}[htpb]
    \centering
    \includegraphics[width=1\textwidth]{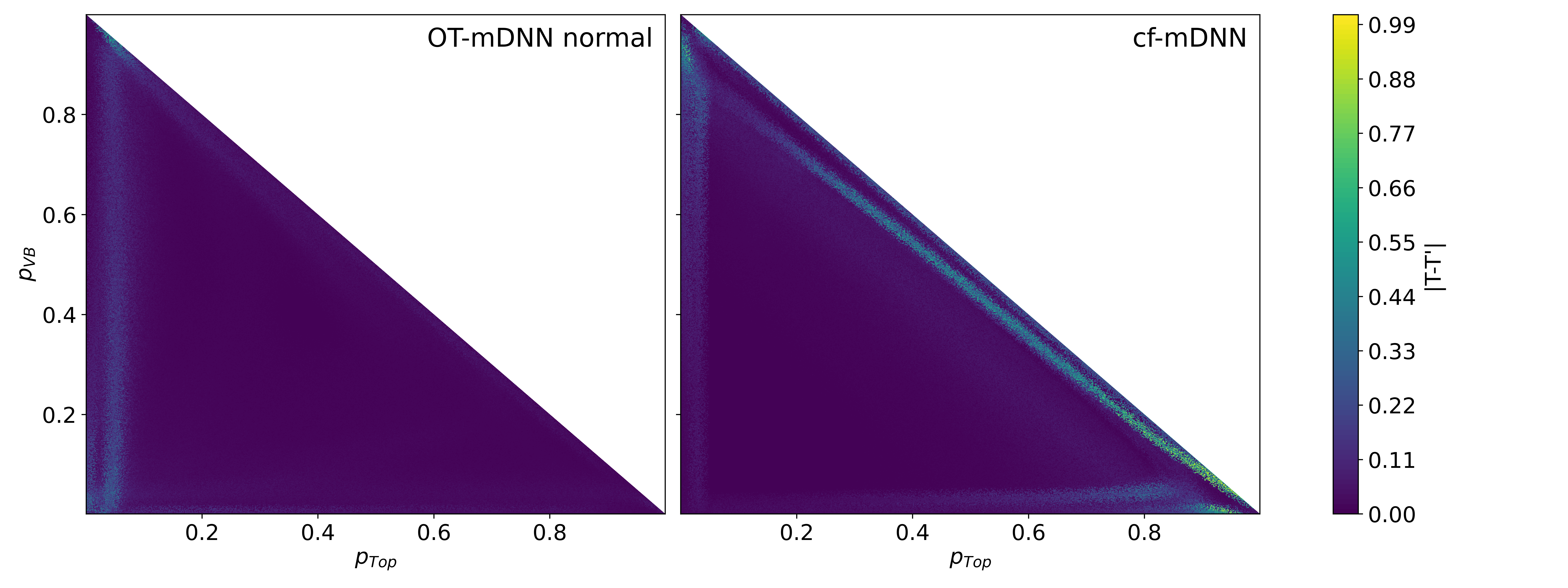}
    \caption{Taking the displacement vectors from Fig.~\ref{fig:displacements_vectors} and calculate the same vectors with a positional uniform variation of $1\%$,
    we can measure the magnitude change of the transport map. The colour indicates the change scaled with $log$.}
    \label{fig:magnitude_difference}
\end{figure*}

OT-mDNN $normal$ has a smooth change in magnitudes indicated by the smooth color transitions in Fig.~\ref{fig:magnitude_difference}. However, there are artifacts within magnitude difference of the cf-mDNN, meaning rapid changes in the transport map. These will break the monotonicity, which can also be seen from the overlapping displacement vectors from Fig.~\ref{fig:displacements_vectors}.
